\documentclass[12pt,preprint]{emulateapj}

\begin{document}
\title{The Bright End of the $z\sim9$ and $z\sim10$ UV Luminosity
  Functions using all five CANDELS Fields\altaffilmark{1}}
\author{R. J. Bouwens\altaffilmark{1}, P. A. Oesch\altaffilmark{2},
  I. Labb{\'e}\altaffilmark{1}, G.D. Illingworth\altaffilmark{3},
  G.G. Fazio\altaffilmark{4}, D. Coe\altaffilmark{5},
  B. Holwerda\altaffilmark{1}, R. Smit\altaffilmark{6},
  M. Stefanon\altaffilmark{1}, P.G. van Dokkum\altaffilmark{2},
  M. Trenti\altaffilmark{7}, M.L.N. Ashby\altaffilmark{4},
  J.-S. Huang\altaffilmark{4}, L. Spitler\altaffilmark{8},
  C. Straatman\altaffilmark{1}, L. Bradley\altaffilmark{5},
  D. Magee\altaffilmark{3}} \altaffiltext{1}{Leiden Observatory,
  Leiden University, NL-2300 RA Leiden, Netherlands}
\altaffiltext{2}{Department of Astronomy, Yale University, New Haven,
  CT 06520} \altaffiltext{3}{UCO/Lick Observatory, University of
  California, Santa Cruz, CA 95064}
\altaffiltext{4}{Harvard-Smithsonian Center for Astrophysics,
  Cambridge, MA, USA} \altaffiltext{5}{Space Telescope Science
  Institute, 3700 San Martin Drive Baltimore, MD 21218}
\altaffiltext{6}{Department of Physics and Astronomy, South Road,
  Durham, DH1 3EE, United Kingdom} \altaffiltext{7}{School of Physics,
  The University of Melbourne, VIC 3010, Australia}
\altaffiltext{8}{Department of Physics \& Astronomy, Macquarie
  University, Sydney, NSW 2109, Australia} \altaffiltext{9}{Based on
  observations made with the NASA/ESA Hubble Space Telescope, which is
  operated by the Association of Universities for Research in
  Astronomy, Inc., under NASA contract NAS 5-26555.}
\begin{abstract}
The deep, wide-area ($\sim$800-900 arcmin$^2$) near-infrared/WFC3/IR +
{\it Spitzer}/IRAC observations over the CANDELS fields have been a
remarkable resource for constraining the bright end of high redshift
UV luminosity functions (LFs).  However, the lack of {\it HST}
1.05$\mu$m observations over the CANDELS fields has made it difficult
to identify $z\sim9$-10 sources robustly, since such data are needed
to confirm the presence of an abrupt Lyman break at 1.2 $\mu$m.  We
report here on the successful identification of many such z$\sim$9-10
sources from a new {\it HST} program (z9-CANDELS) that targets the
highest-probability z$\sim$9-10 galaxy candidates with observations at
1.05$\mu$m, to search for a robust Lyman-break at 1.2$\mu$m.  The
potential z$\sim$9-10 candidates are preselected from the full {\it
  HST}, {\it Spitzer}/IRAC S-CANDELS observations, and the
deepest-available ground-based optical+near-infrared observations
(CFHTLS-DEEP+HUGS+UltraVISTA+ZFOURGE). We identified 15 credible
z$\sim$9-10 galaxies over the CANDELS fields.  Nine of these galaxies
lie at z$\sim$9 and 5 are new identifications.  Our targeted follow-up
strategy has proven to be very efficient in making use of scarce {\it
  HST} time to secure a reliable sample of $z\sim9$-10 galaxies.
Through extensive simulations, we replicate the selection process for
our sample (both the preselection and follow-up) and use it to improve
current estimates for the volume density of bright z$\sim$9 and
z$\sim$10 galaxies.  The volume densities we find are
5$_{-2}^{+3}\times$ and 8$_{-3}^{+9}\times$ lower, respectively, than
found at $z\sim 8$.  When compared with the best-fit evolution (i.e.,
$d\log_{10}\rho_{UV}/dz=-0.29\pm0.02$) in the $UV$ luminosities
densities from $z\sim 8$ to $z\sim4$ integrated to $0.3L_{z=3}^{*}$
($-20$ mag), these luminosity densities are $2.6_{-0.9}^{+1.5}\times$
and $2.2_{-1.1}^{+2.0}\times$ lower, respectively, than the
extrapolated trends.  Our new results are broadly consistent with the
``accelerated evolution'' scenario at $z>8$, as seen in many
theoretical models.
\end{abstract}

\section{Introduction}

The first galaxies are thought to have formed in the first 300-400 Myr
of the universe.  Over the last few years, remarkable progress has
been made in extending samples back to this time, with more than
$\sim$700 probable galaxies identified at $z\gtrsim6.3$ with {\it HST}
(Bouwens et al.\ 2015; McLure et al.\ 2013; Finkelstein et al.\ 2015)
and 20-30 candidate galaxies identified as far back as redshifts
$z\sim 9$-11 (Bouwens et al.\ 2011a, 2014b; Zheng et al.\ 2012; Coe et
al.\ 2013; Ellis et al.\ 2013; McLure et al.\ 2013; Oesch et
al.\ 2013, 2014, 2015; Zitrin et al.\ 2014; Ishigaki et al.\ 2015;
Bouwens et al.\ 2015; McLeod et al.\ 2015).

At present and over the next year, considerable resources are being
devoted to both the discovery and study of ultra-faint galaxies with
{\it HST} from the new Frontier Fields initiative (e.g., Lotz et
al.\ 2014; Coe et
al.\ 2015)\footnote{http://www.stsci.edu/hst/campaigns/frontier-fields/},
The goal of this initiative is to combine the power of gravitational
lensing from galaxy clusters with very deep exposures with the {\it
  Hubble} and {\it Spitzer} Space Telescopes.  840 orbits of {\it HST}
observations are being invested in deep optical/ACS + near-IR/WFC3/IR
observations of 6 galaxy clusters.  Deep observations of a ``blank''
field outside the galaxy clusters are also being obtained in parallel
with observations over the clusters.

Despite the considerable focus by the community on the Hubble Frontier
Fields observations over galaxy clusters and deep fields (e.g., Atek
et al.\ 2014, 2015; Zheng et al.\ 2014; Coe et al.\ 2015; Ishigaki et
al.\ 2015; Oesch et al.\ 2015), it is also possible to uncover modest
numbers of luminous $z\sim9$-10 galaxies over wide-field surveys as
Oesch et al.\ (2014) first illustrated through the identification of
six intrinsically-luminous $z\sim 9$-10 candidate galaxies over the
GOODS-North and GOODS-South CANDELS fields (Grogin et al.\ 2011;
Koekemoer et al.\ 2011).  These sources allowed us to set some initial
constraints on the rate at which $UV$-luminous galaxies evolve with
cosmic time, as well as providing some constraints on the approximate
shape of the $UV$ LF at $z=9$-10.

Due to the inherent brightness of such sources, these sources are also
valuable for efforts to measure the physical properties of galaxies at
very early times.  Measurements of the $UV$-continuum slopes (Oesch et
al.\ 2014; Wilkins et al.\ 2016), Balmer-break amplitudes (Oesch et
al.\ 2014), stellar masses (Oesch et al.\ 2014), and sizes (Holwerda
et al.\ 2014; Shibuya et al.\ 2015) can all be achieved using very
bright galaxies.

\begin{deluxetable*}{lcllcllcl}
\tablewidth{0pt}
\tablecolumns{9}
\tabletypesize{\footnotesize}
\tablecaption{Observational Data Used to Identify\tablenotemark{$\dagger$} the Bright $z\sim9$-10 Candidate Galaxies over the CANDELS UDS, COSMOS, and EGS fields.\tablenotemark{*}\label{tab:dataset}}
\startdata
\tableline \\
\multicolumn{9}{c}{Two-Part Search Strategy (Preselection + Follow-up: \S3, \S4)}\\\\
\multicolumn{3}{l}{CANDELS UDS} & \multicolumn{3}{l}{CANDELS COSMOS} & \multicolumn{3}{l}{CANDELS EGS} \\
\tableline \\
       & 5$\sigma$ &    &        & 5$\sigma$ &        &        & 5$\sigma$ &  \\
Filter\tablenotemark{$\dagger$} & Depth\tablenotemark{a} & Source & Filter\tablenotemark{$\dagger$} & Depth\tablenotemark{a}     & Source & Filter\tablenotemark{$\dagger$} & Depth\tablenotemark{a}     & Source \\
\tableline \\
$V_{606}$ & 26.8 & {\it HST}/ACS & $V_{606}$ & 26.5 & {\it HST}/ACS & $V_{606}$ & 27.3 & {\it HST}/ACS \\
$I_{814}$ & 26.8 & {\it HST}/ACS & $I_{814}$ & 26.5 & {\it HST}/ACS & $I_{814}$ & 27.1 & {\it HST}/ACS \\
$J_{125}$ & 26.3 & {\it HST}/WFC3 & $J_{125}$ & 26.1 & {\it HST}/WFC3 & $J_{125}$ & 26.4 & {\it HST}/WFC3 \\
$JH_{140}$ & 26.1 & {\it HST}/WFC3 & $JH_{140}$ & 25.8 & {\it HST}/WFC3 & $JH_{140}$ & 25.6 & {\it HST}/WFC3 \\
$H_{160}$ & 26.5 & {\it HST}/WFC3 & $H_{160}$ & 26.3 & {\it HST}/WFC3 & $H_{160}$ & 26.6 & {\it HST}/WFC3 \\
$u$ & 25.8 & CFHT/Megacam & $u$ & 27.7 & CFHT/Megacam & $u$ & 27.4 & CFHT/Megacam \\
$B$ & 28.0 & Subaru/Suprime-Cam & $B+g$ & 28.4 & Subaru/Suprime-Cam + & $g$ & 27.8 & CFHT/Megacam \\
$V+r$ & 28.0 & Subaru/Suprime-Cam &       & & CFHT/Megacam & $r$ & 27.6 & CFHT/Megacam \\
$i$ & 27.4 & Subaru/Suprime-Cam   & $V+r$ & 27.9 & Subaru/Suprime-Cam + & $i+y$ & 27.4 & CFHT/Megacam \\
$z$ & 26.3 & Subaru/Suprime-Cam   &       & & CFHT/Megacam & $z$ & 26.0 & CFHT/Megacam \\
$Y$ & 25.9 & VLT/HAWKI/HUGS       & $i+y$ & 27.7 & Subaru/Suprime-Cam + & $K$ & 24.1 & UKIRT/WIRCam \\
$J_1$ & 25.6 & Magellan/FOURSTAR  &       & & CFHT/Megacam         & 3.6$\mu$m & 25.4 & {\it Spitzer}/S-CANDELS \\
$J_2$ & 25.7 & Magellan/FOURSTAR  & $z$   & 26.4 & Subaru/Suprime-Cam + & 4.5$\mu$m & 25.3 & {\it Spitzer}/S-CANDELS \\
$J$ & 25.4 & UKIRT/WFCAM          &       & & CFHT/Megacam \\
$J_3$ & 25.4 & Magellan/FOURSTAR  & $Y$   & 26.1 & UltraVISTA \\
$H$ & 24.6 & UKIRT/WFCAM          & $J_1$ & 25.6 & Magellan/FOURSTAR \\
$H_s$ & 25.0 & Magellan/FOURSTAR  & $J_2$ & 25.5 & Magellan/FOURSTAR \\
$H_l$ & 24.8 & Magellan/FOURSTAR  & $J$   & 25.3 & UltraVISTA \\
$K_s$ & 25.5 & VLT/HAWKI/HUGS +   & $J_3$ & 25.3 & Magellan/FOURSTAR \\
      & & UKIRT/WFCAM +      & $H_s$ & 24.7 & Magellan/FOURSTAR \\
      & & Magellan/FOURSTAR     & $H$   & 25.0 & UltraVISTA     \\ 
3.6$\mu$m & 25.4 & {\it Spitzer}/S-CANDELS & $H_l$ & 24.7 & Magellan/FOURSTAR \\
4.5$\mu$m & 25.4 & {\it Spitzer}/S-CANDELS & $K_s$ & 25.3 & UltraVISTA + \\
          & &                   & & & Magellan/FOURSTAR \\
 & & & 3.6$\mu$m & 25.3 & {\it Spitzer}/S-CANDELS \\
 & & & 4.5$\mu$m & 25.3 & {\it Spitzer}/S-CANDELS \\
\tableline \\ \\
\multicolumn{9}{c}{Direct Search Strategy for $z\geq8.4$ Galaxies (\S5)}\\\\
\multicolumn{3}{l}{CANDELS GOODS-South} & \multicolumn{3}{l}{ERS} & \multicolumn{3}{l}{CANDELS GOODS-North} \\
\tableline \\
$B_{435}$ & 27.1-27.3 & {\it HST}/ACS & $B_{435}$ & 27.1 & {\it HST}/ACS & $B_{435}$ & 27.2-27.3 & {\it HST}/ACS \\
$V_{606}$ & 27.4-27.7 & {\it HST}/ACS & $V_{606}$ & 27.4 & {\it HST}/ACS & $V_{606}$ & 27.4 & {\it HST}/ACS \\
$i_{775}+$ &  &  & $i_{775}+$ &  &  & $i_{775}+$ &  \\
$I_{814}$ & 27.5-27.6 & {\it HST}/ACS & $I_{814}$ & 27.3 & {\it HST}/ACS & $I_{814}$ & 27.2-27.7 & {\it HST}/ACS \\
$z_{850}$ & 26.8-26.9 & {\it HST}/ACS & $z_{850}$ & 26.7 & {\it HST}/ACS & $z_{850}$ & 26.9-27.0 & {\it HST}/ACS \\
$Y_{105}$ & 26.4-27.0 & {\it HST}/WFC3 & $Y_{098}$ & 26.5 & {\it HST}/WFC3 & $Y_{105}$ & 26.5-26.8 & {\it HST}/WFC3 \\
$J_{125}$ & 26.5-27.0 & {\it HST}/WFC3 & $J_{125}$ & 27.0 & {\it HST}/WFC3 & $J_{125}$ & 26.4-27.2 & {\it HST}/WFC3 \\
$JH_{140}$ & 26.1 & {\it HST}/WFC3 & $JH_{140}$ & 25.8 & {\it HST}/WFC3 & $JH_{140}$ & 25.6 & {\it HST}/WFC3 \\
$H_{160}$ & 26.5-27.0 & {\it HST}/WFC3 & $H_{160}$ & 26.9 & {\it HST}/WFC3 & $H_{160}$ & 26.5-27.1 & {\it HST}/WFC3 \\
$K_s$ & 26.5 & VLT/HAWKI/HUGS +   & $K_s$ & 26.5 & VLT/HAWKI/HUGS + \\
      & & VLT/ISAAC +  &       & & VLT/ISAAC +  \\
      & & PANIC +      &       & & PANIC +     \\
      & & Magellan/FOURSTAR      &       & & Magellan/FOURSTAR     \\
3.6$\mu$m & 25.8 & {\it Spitzer}/S-CANDELS & 3.6$\mu$m & 25.8 & Spitzer/S-CANDELS & 3.6$\mu$m & 25.8 & {\it Spitzer}/S-CANDELS \\
4.5$\mu$m & 25.8 & {\it Spitzer}/S-CANDELS & 4.5$\mu$m & 25.8 & Spitzer/S-CANDELS & 4.5$\mu$m & 25.8 & {\it Spitzer}/S-CANDELS 
\enddata
\tablenotetext{$\dagger$}{For each source in our search fields, flux
  measurements are derived based on the all the observational data
  presented in this table.  All of these measurements are used in
  deriving a redshift likelihood distribution for individual
  sources.}
\tablenotetext{a}{The $5\sigma$ depth are estimated from the median
  $5\sigma$ uncertainties on the total flux measurements of sources
  found over our search fields with $H_{160,AB}$-band magnitudes of
  26.0-26.5.}
\end{deluxetable*}

Despite the usefulness of bright $z\sim 9$-10 galaxies for addressing
many contemporary science questions, current samples of these objects
remain quite small and constraints on their volume densities poor.  Of
particular note, the recent Oesch et al.\ (2014) sample over the
GOODS-North and GOODS-South fields only contained 2 bright $z\sim9$
and 4 bright $z\sim10$ galaxies.  With such small samples, current
uncertainties on the volume density of bright $z\sim9$-10 galaxies are
large indeed ($\gtrsim0.3$-0.4 dex).  This is especially the case when
one considers the impact of field-to-field variations (``cosmic
variance'') which is as large as a factor of two across the CANDELS
fields, e.g., see Figure 14 from Bouwens et al.\ (2015), and may be
even larger for the brightest sources (Bowler et al.\ 2015;
Roberts-Borsani et al.\ 2016).  Clearly, we require many independent
lines of sight on the $z\sim9$-10 universe to average over the
large-scale structure. Unfortunately, the Frontier Fields Initiative
will not significantly help with this issue for luminous sources,
given the limited area covered by observations from this program.

Nevertheless, there is a huge quantity of {\it HST} and {\it Spitzer}
data already available that can be used to construct larger samples of
bright $z\sim 9$-10 galaxies.  The most significant of these data sets
are the $\sim$500 arcmin$^2$ CANDELS UDS, COSMOS, and EGS fields which
feature very deep optical, near-IR, and {\it Spitzer}/IRAC
observations.  These observations are very useful for the robust
detection of bright $z\sim9$-10 candidates and also confirming a blue
color redward of the break, distinguishing such galaxies from dusty,
red galaxies at $z\sim1$-3.  While possessing great potential, the
CANDELS UDS, COSMOS, and EGS search areas lack correspondingly deep
observations at $1.05\mu$m, just blueward of the Lyman-break in
candidate $z\sim9$-10 galaxies which is important for confirming a
spectral break at $\sim$1.2$\mu$m and also distinguishing these
$z\gtrsim9$ galaxy candidates from Balmer-break sources at $z\sim1$-3.

\begin{figure}
\epsscale{1.15} \plotone{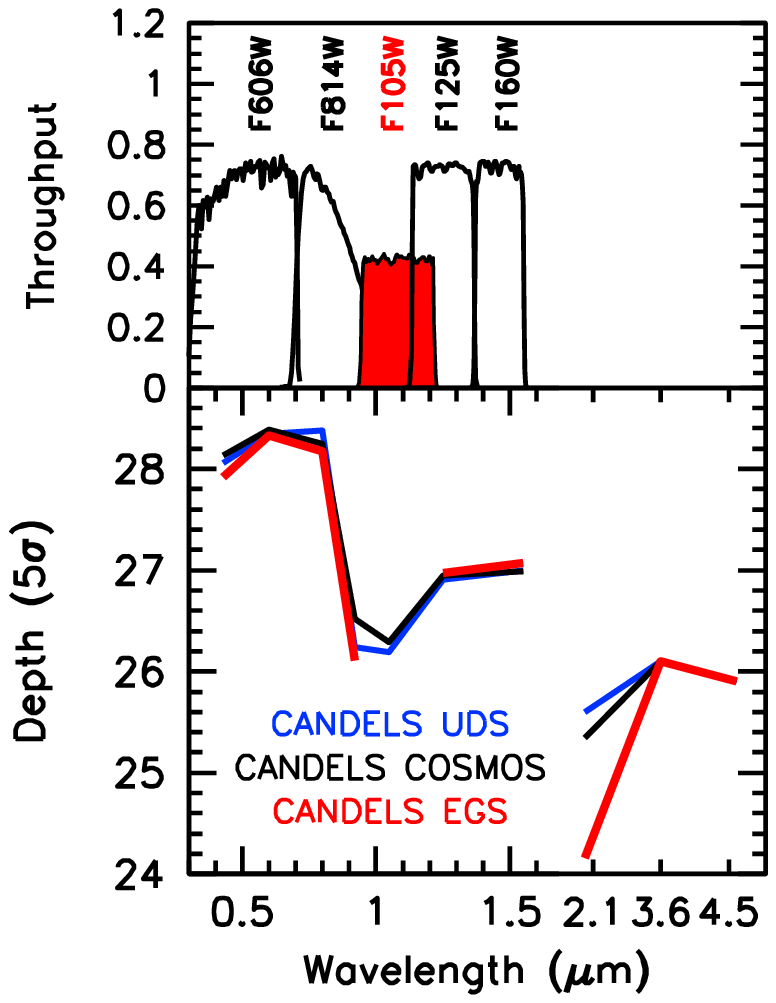}
\caption{(\textit{upper panel}) Wavelength sensitivity curves for the
  filters (F606W, F814W, F125W, F160W) in which deep {\it HST} observations
  are available over the CANDELS UDS, COSMOS, and EGS fields
  (\textit{black}), as well as those (F105W) primarily obtained by our
  follow-up program z9-CANDELS (\textit{red}).  (\textit{lower panel})
  $5\sigma$ median depths of the observations versus wavelength
  available over the CANDELS UDS, COSMOS, and EGS fields (see also the
  compilations in Table 1 and Figure 2 of Bouwens et al.\ 2015).  The
  depths plotted here are binned in such a way to combine all the data
  (Table~\ref{tab:dataset}) that exist in 0.1-0.15$\mu$m segments.
  The depths do not include the ZFOURGE observations here, since those
  observations only cover $\gtrsim$65\% of each CANDELS field (adding
  some useful depth from $1.0\mu$m to $1.7\mu$m).  There is a modest
  wavelength gap between the deeper observations at
  $\sim0.6$-0.9$\mu$m and that which exists at $\sim1.2$-1.6$\mu$m.
  While some ($\sim$26-mag, $5\sigma$) observations exist at
  $1.05\mu$m to probe below the putative Lyman-break for $z\sim9$-10
  galaxy candidates, the addition of deep observations at $1\mu$m with
  {\it HST} can significantly improve current constraints on the robustness
  of the Lyman-break at $1\mu$m.\label{fig:b9fig}}
\end{figure}

Fortunately, we can overcome the aforementioned limitations of the
CANDELS UDS, COSMOS, and EGS data sets by leveraging essentially all
existing observations over these fields (Table~\ref{tab:dataset}) to
first identify the highest probability $z\sim9$-10 candidates over
these fields and then obtaining targeted follow-up observations of
these candidates at $1.05\mu$m to determine which are likely at $z>8$
(Figure~\ref{fig:b9fig}).  In cycle 21, we successfully proposed for
such a follow-up program of plausible candidate $z\sim9$-10 galaxies
over the CANDELS UDS, COSMOS, and EGS fields.  Observations from this
program -- which we call z9 (redshift 9)-CANDELS (Bouwens 2014: GO
13792) are now complete and cover all 12 of the primary candidates
from that program.  Based on the information we obtain from our
proposed follow-up observations and the selection criteria we use in
identifying our initial sample of 12 candidate $z\sim9$-10 galaxies
from these three CANDELS fields, we can derive constrants on the
volume density of luminous $z\sim9$-10 galaxies.  Searches over
CANDELS-GOODS-North, CANDELS-GOODS-South, the ERS fields can be
further used to improve the constraints we can obtain on the bright
end of the $z\sim9$-10 LFs.

In this paper, we describe (1) the preselection we utilized to
identify candidate $z\sim9$-10 galaxies from the CANDELS-UDS, COSMOS,
and EGS fields for {\it HST} follow-up observations and (2) the
results from this program.  Our primary scientific objective is in
obtaining the best available constraints on the volume density of
especially luminous $z\sim9$ and $z\sim10$ galaxies.  Through such
constraints, we have a direct measure of how fast (1) the bright end
of the $UV$ LF and (2) $UV$-luminous galaxies evolve.  Through
comparison with the volume density of fainter sources, the present
search results also allow us to constrain the overall shape of the
$UV$ LF.  Finally, we would expect our selection to allow us to
considerably expand the overall sample of bright $z\sim9$ and
$z\sim10$ galaxies available over the CANDELS fields.  This has value
both for the further characterization of the physical properties of
$z\gtrsim9$ galaxies and as bright sources to target with early JWST
observations.  These bright samples will be further enhanced with
bright $z\sim9$-10 galaxies from the BoRG$_{[z910]}$ program (Trenti
2014).

Here is a brief plan for this paper.  In \S2, we include a description
of the observational data that we utilize to identify high-probability
$z\sim9$-10 galaxies over the CANDELS-UDS, COSMOS, and EGS fields.  In
\S3, we describe our criteria for performing photometry and
identifying high-probability $z\sim9$-10 galaxy candidates over the
CANDELS-UDS, COSMOS, and EGS fields.  In \S4, we describe the results
of the z9-CANDELS program where we use these observations to ascertain
the likely nature of our selected $z\sim9$-10 candidate galaxies.  In
\S5, we describe our search results for bright $z\gtrsim8.4$ over the
CANDELS GOODS-North, GOODS-South fields, and ERS fields, extending
previous work by Oesch et al.\ (2014: see also McLure et al.\ 2013 who
also conducted such a search over the GOODS-South field).  Finally, in
\S6, we make use of these search results to provide the first
constraints on the bright end of the $z\sim9$ and $z\sim10$ LFs using
a search over all five CANDELS fields.

For consistency with previous work, we quote results in terms of the
luminosity $L_{z=3}^{*}$ Steidel et al.\ (1999) derived at $z\sim3$,
i.e., $M_{1700,AB}=-21.07$.  We refer to the {\it HST} F435W, F606W, F775W,
F814W, F850LP, F098M F105W, F125W, F140W, and F160W bands as
$B_{435}$, $V_{606}$, $i_{775}$, $I_{814}$, $z_{850}$, $Y_{098}$,
$Y_{105}$, $J_{125}$, $JH_{140}$, and $H_{160}$, respectively, for
simplicity.  Where necessary, we assume $\Omega_0 = 0.3$,
$\Omega_{\Lambda} = 0.7$, and $H_0 = 70\,\textrm{km/s/Mpc}$.  All
magnitudes are in the AB system (Oke \& Gunn 1983).

\section{Observational Data}

In the present analysis, we conduct a search for bright $z\sim9$-10
candidate galaxies over the $\sim$450 arcmin$^2$ region within the
CANDELS-UDS, COSMOS, and EGS fields with the deepest {\it HST} optical/ACS
and near-IR/WFC/IR observations ($\sim$75-80\% of the WFC3/IR area).

In conducting this search, we utilize the reductions of the {\it HST}
observations described in Bouwens et al.\ (2015).  Those reductions
include all observations associated with the AEGIS, COSMOS, and
CANDELS {\it HST} surveys and SNe follow-up programs, including the
$JH_{140}$-band observations associated with 3D-{\it HST} (Brammer et
al.\ 2012) and AGHAST (Weiner et al.\ 2014) programs.

Beyond the {\it HST} observations themselves, perhaps the most
valuable data set that we can leverage in our search for probable
$z\sim9$-10 galaxies is the very deep {\it Spitzer}/IRAC S-CANDELS
observations over the CANDELS fields (Ashby et al.\ 2015), which when
combined with {\it Spitzer}/IRAC SEDS observations (Ashby et
al.\ 2013) reach 50 hours in depth (26.0 mag at $5\sigma$:
2$''$-diameter apertures).  Those observations provide us with
constraints on the spectral slope of galaxies redward of the $H_{160}$
band, which when combined with evidence for a break across the
$J_{125}$ and $H_{160}$ bands and a non-detection at optical
wavelengths is strongly suggestive of a $z\sim9$-10 galaxy.

\begin{figure*}
\epsscale{1.04} \plotone{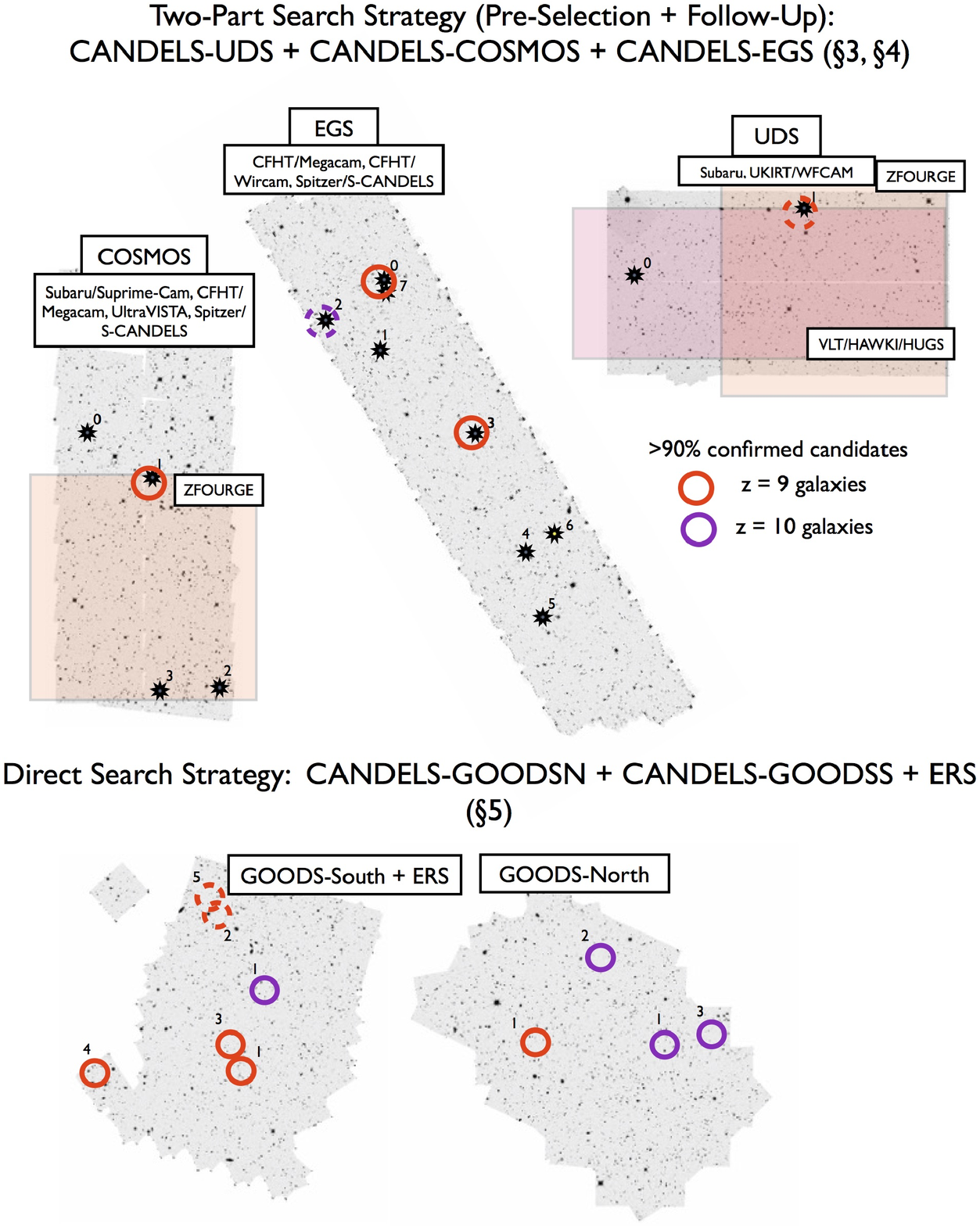}
\caption{$H_{160}$-band images of the CANDELS-UDS, COSMOS, and EGS
  search data that we used to identify tentative candidate $z\sim9$-10
  galaxies.  The positions of the candidate $z\sim9$-10 sources are
  indicated by the stars on these mosaics.  The numbers adjacent to
  the stars indicate the identity of the tentative $z\sim9$-10
  candidate (identical the numbering scheme employed in
  Table~\ref{tab:catalog}).  Those stars shaded in grey indicate
  sources that were explicitly preselected for targeted follow-up
  observations from our 11-orbit z9-CANDELS program, while those
  shaded in yellow were not preselected and only incidentally targeted
  (lowest two rows in Table~\ref{tab:catalog}).  Also shown are the
  regions within these fields where deep near-IR observations are
  available from programs like HUGS (Fontana et al.\ 2014) or ZFOURGE
  (Labb{\'e} et al.\ 2015, in prep) which cover most but not all of
  the area over the targeted CANDELS fields.  The candidates enclosed
  in red or purple circles appear very likely ($>$90\% confidence) to
  be at $z\sim9$ or $z\sim10$, respectively, based on the available
  photometric constraints (obtained with our {\it HST} follow-up program or
  archival observations).\label{fig:layout}}
\end{figure*}

In addition, we also make use of all significant, public ground-based
observations over these fields, including optical observations from
Subaru Suprime-Cam [CANDELS-COSMOS; CANDELS-UDS] and CFHT/Megacam
[CANDELS-COSMOS; CANDELS-EGS] and deep near-IR observations from VISTA
[CANDELS-COSMOS], UKIRT/WFCAM [CANDELS-UDS], and VLT/HAWKI
[CANDELS-UDS], Magellan/FOURSTAR [CANDELS-COSMOS; CANDELS-UDS], and
CFHT/WIRCam [CANDELS-EGS].  The deep optical observations allow us to
search for faint optical flux in the $z\sim9$-10 candidate identified
over the CANDELS-UDS/COSMOS/EGS fields, while the near-IR observations
allow us to test for the presence of a putative break at $1.2\mu$m,
verify that candidates show no flux blueward of the break, and test
for a flat $UV$-continuum redward of the break.

In our analysis of data over the five CANDELS fields and $\sim$40
arcmin$^2$ ERS field (Windhorst et al.\ 2011), we use the Bouwens et
al.\ (2015) reductions of the {\it HST} observations over all five
CANDELS fields, the version 7 reduction of the deep CFHT legacy survey
observations over the COSMOS and EGS
fields,\footnote{http://www.cfht.hawaii.edu/Science/CFHTLS} the public
v2.0 reductions of the UltraVISTA observations (McCracken et
al.\ 2012), the Cirasuolo et al.\ (2010) redutions of the deep Subaru
Suprime-Cam observations over the UDS/SXDS field (Furusawa et
al. 2008), the Bouwens et al.\ (2015) reductions of the HUGS HAWK-I
observations (Fontana et al.\ 2014), the public reductions of the
WIRCam deep survey $K_s$-band observations over the CANDELS EGS field
(McCracken et al.\ 2010; Bielby et al.\ 2012), the v0.9.3/v0.95.5
reductions of the ZFOURGE COSMOS/UDS observations (I. Labb{\'e} et
al.\ 2015, in prep), the IUDF reductions of the {\it Spitzer}/IRAC
observations over the GOODS-South and GOODS-North fields (Labb{\'e} et
al.\ 2015), and the public reductions of the {\it Spitzer} SEDS and
S-CANDELS programs (Ashby et al.\ 2013; Ashby et al.\ 2015).

Table~\ref{tab:dataset} provides a convenient summary of all the
observational data we utilize.  Combining the flux measurements from
the different data sets, the $5\sigma$ depths of these fields (derived
from the median uncertainties on the total flux measurements) range
from $\sim$28 mag at $<$0.8$\mu$m, $\sim$26.0-26.5 mag at
$\sim$0.9$\mu$m, $\sim$26.0 mag at $1.05\mu$m, $\sim$26.6 mag at
$\sim$1.2-1.6$\mu$m, 24.1-25.5 mag at $2.3\mu$m, and 25.6-25.9 mag at
3.6$\mu$m + 4.5$\mu$m.  We refer the interested reader to Figure 3
from Bouwens et al.\ (2015) for a graphical representation of these
depths as a function of wavelength.

\begin{deluxetable*}{ccc}
\tabletypesize{\footnotesize}
\tablecaption{Selection Criteria Utilized in Assembling our $z\sim9$ and $z\sim10$ Samples\label{tab:selcrit}}
\tablehead{
\colhead{Redshift}            & \multicolumn{2}{c}{Selection Criteria} \\
\colhead{Sample}     & \colhead{Preselection for Targeted {\it HST} Follow-up} & \colhead{After {\it HST} Follow-up}} \\
\multicolumn{3}{c}{CANDELS-UDS + CANDELS-COSMOS}\\
9 & $(J_{125}-H_{160}>0.5)\wedge(H_{160}-[3.6]<1.4)\wedge$  (S/N in both $V_{606}$ and $I_{814}$ $< 2$)$\wedge$ & $(P(z>8)>0.9)$$\wedge$$(8.4<z_{phot}<9.5)$ \\
  & (rms S/N in $V_{606}$ and $I_{814}$ $< 1)\wedge$ (S/N($H_{160}$)$>5)\wedge(\chi_{JH_{140}H_{160}} ^2)>36)$$\wedge$ &  \\
  & ($\chi_{K,[3.6],[4.5]} ^2 >2) \wedge (P_{pre} (z>8)>0.5)\wedge(P_{post} (z>8)>0.9)$ \\\\
10 & idem & $(P(z>8)>0.9)\wedge(9.5<z_{phot}<11)$    \\\\
\multicolumn{3}{c}{CANDELS-EGS}\\
9 & $(J_{125}-H_{160}>0.5)\wedge(H_{160}-[3.6]<1.4)\wedge$  (S/N in both $V_{606}$ and $I_{814}$ $< 2$)$\wedge$ & $(P(z>8)>0.9)$$\wedge$$(8.4<z_{phot}<9.5)$ \\
  & (rms S/N in $V_{606}$ and $I_{814}$ $< 1)\wedge$ (S/N($H_{160}$)$>5)\wedge(\chi_{JH_{140}H_{160}})>6)$$\wedge$ &  \\
  & ($\chi_{K,[3.6],[4.5]} ^2 >2)\wedge(P_{post} (z>8)>0.9)$ \\\\
10 & idem & $(P(z>8)>0.9)\wedge(9.5<z_{phot}<11)$    \\\\
\multicolumn{3}{c}{CANDELS-GOODS-North + CANDELS-GOODS-South + ERS}\\
9 & \multicolumn{2}{c}{((Y-dropout criterion from Bouwens+2015) $\vee (J_{125}-H_{160}>0.5))\wedge(H_{160}-[3.6]<1.4)\wedge$}\\
  & \multicolumn{2}{c}{(S/N in both $V_{606}$ and $I_{814}$ $< 2)\wedge(\chi_{B_{435}V_{606}i_{775}I_{814}z_{850}} ^2 < 4)\wedge$}\\
  & \multicolumn{2}{c}{$(P(z>8)>0.8)$$\wedge$($8.4<z_{phot}<9.5$)}  \\\\
10 & \multicolumn{2}{c}{((Y-dropout criterion from Bouwens+2015) $\vee (J_{125}-H_{160}>0.5))\wedge(H_{160}-[3.6]<1.4)\wedge$}\\
  & \multicolumn{2}{c}{(S/N in both $V_{606}$ and $I_{814}$ $< 2)\wedge(\chi_{B_{435}V_{606}i_{775}I_{814}z_{850}} ^2 < 4)\wedge$}\\
  & \multicolumn{2}{c}{$(P(z>8)>0.8)$$\wedge$($9.5<z_{phot}<11.0$)}  
\enddata
\tablenotetext{*}{Redshift likelihood probability $P(z)$ are computed using our flux meaurements in all photometric bands listed in Table~\ref{tab:dataset}.  $P_{pre}(z>8)$ indicates the probability that a source has a redshift greater than 8 before acquiring any follow-up observations, while $P_{post}(z>8)$ indicates the probability that a source has a redshift greater than 8, after obtaining the 1-orbit of follow-up {\it HST} observations (assuming the follow-up observations yielded a measured flux of 0$\pm$12 nJy in the $Y_{105}$-band filter).}
\end{deluxetable*}

\begin{deluxetable*}{cccccccc}
\tablewidth{0pt}
\tablecolumns{7}
\tabletypesize{\footnotesize}
\tablecaption{$z\sim9$-10 Candidate Galaxies over the CANDELS UDS, COSMOS, and EGS program targeted with our z9-CANDELS follow-up program.\label{tab:catalog}}
\tablehead{
\colhead{ID} & \colhead{R.A.} & \colhead{Dec} & \colhead{$H_{160,AB}$} & \colhead{$z_{phot,pre}$\tablenotemark{a}} & \colhead{P$_{pre}$($z>8$)\tablenotemark{a}} & \colhead{$z_{phot,post}$\tablenotemark{b}} & \colhead{P$_{post}$ ($z>8$)\tablenotemark{b}}}
\startdata
\multicolumn{8}{c}{$z=9$-10 Candidates Preselected for Targeted Follow-Up Observations with {\it HST}}\\
COS910-0 & 10:00:43.16 & 02:25:10.5 & 26.2$\pm$0.1 & 9.1 & 0.72 & 7.8 & 0.47 \\
COS910-1 & 10:00:30.34 & 02:23:01.6 & 26.4$\pm$0.2 & 9.0 & 0.95 & 9.0 & 0.99 \\
COS910-2 & 10:00:14.91 & 02:12:10.8 & 26.3$\pm$0.2 & 9.3 & 0.74 & 9.3 & 0.37 \\
COS910-3 & 10:00:27.98 & 02:11:49.5 & 25.9$\pm$0.1 & 9.2 & 0.63 & 2.3 & 0.27 \\
UDS910-0 & 02:17:55.50 & $-$05:11:41.3 & 26.4$\pm$0.2 & 8.8 & 0.72 & 1.7 & 0.09 \\
UDS910-1 & 02:17:21.96 & $-$05:08:14.7 & 26.6$\pm$0.2 & 8.7 & 0.74 & 8.6 & 0.74 \\
EGS910-0 & 14:20:23.47 & 53:01:30.5 & 26.2$\pm$0.1 & 9.1 & 0.67 & 9.1 & 0.92 \\
EGS910-1 & 14:20:21.54 & 52:57:58.4 & 26.6$\pm$0.1 & 8.9 & 0.19\tablenotemark{$\dagger$} & 0.4 & 0.02 \\
EGS910-2 & 14:20:44.31 & 52:58:54.4 & 26.7$\pm$0.2 & 9.6 & 0.69 & 9.6 & 0.71 \\
EGS910-3 & 14:19:45.28 & 52:54:42.5 & 26.4$\pm$0.2 & 8.9 & 0.64 & 9.0 & 0.97 \\
EGS910-4 & 14:19:23.59 & 52:49:23.4 & 26.2$\pm$0.2 & 9.2 & 0.10\tablenotemark{$\dagger$} & 1.0 & 0.02 \\
EGS910-5 & 14:19:11.08 & 52:46:25.7 & 25.8$\pm$0.1 & 9.2 & 0.28 & 1.8 & 0.11 \\
\\
\multicolumn{8}{c}{$z\sim9$-10 Galaxy Candidates Targeted at No Additional Cost (Not Preselected)\tablenotemark{c}}\\
EGS910-6 & 14:19:13.84 & 52:50:44.7 & 26.6$\pm$0.2 & 9.3 & 0.40 & 7.0 & 0.00 \\
EGS910-7 & 14:20:23.72 & 53:01:38.3 & 26.0$\pm$0.1 & --- & ---- & 2.4 & 0.18
\enddata
\tablenotetext{a}{Best-fit $z>4$ redshift and integrated $z>8$ likelihood for source derived from our {\it HST}+{\it Spitzer}/IRAC+ground-based photometry (Table~\ref{tab:dataset}), before obtaining observations from our z9-CANDELS follow-up program.}
\tablenotetext{b}{Best-fit redshift and integrated $z>8$ likelihood for source derived from our {\it HST}+{\it Spitzer}/IRAC+ground-based photometry (Table~\ref{tab:dataset}), after obtaining observations from our z9-CANDELS follow-up program.}
\tablenotetext{c}{These sources could be fit within the same WFC3/IR tiles, as our primary targets, and hence required no additional {\it HST} time to investigate.}
\tablenotetext{$\dagger$}{Over the CANDELS EGS field, we selected sources which, if they showed  a null detection in the $Y_{105}$ band in a 1-orbit integration, could be confirmed with $>$90\% probability to lie at $z>8$.  While these two sources initially only showed a modest probability for being at $z>8$, their SEDs were nevertheless consistent with lying at $z>8$ (particularly if a null detection at $1.05\mu$m could be confirmed).}
\end{deluxetable*}

\section{$z\sim9$-10 Selection}

\subsection{Catalog Construction and Photometry}

As in previous work (e.g., Bouwens et al.\ 2007, 2011b, 2015), we use
a modified version of the SExtractor software (Bertin \& Arnouts 1996)
for constructing our {\it HST} source catalogs that lie at the core of the
$z\sim9$-10 selection we perform.  SExtractor is run in dual mode,
with source detection done off the $H_{160}$-band images and
photometry performed on the $V_{606}$, $I_{814}$, $J_{125}$,
$JH_{140}$, and $H_{160}$ images one at a time.  Color measurements
are made in small scalable apertures using Kron-style (1980)
photometry and a Kron-parameter of 1.2.  Fluxes measured in these
small scalable aperture are then corrected to total in two steps.  In
the first step, we multiply each of the fluxes by the excess flux seen
in the larger scalable apertures (Kron parameter of 2.5) for the
$H_{160}$-band over that present in smaller scalable apertures.  In
the second step, we correct for the light on the wings of the PSF and
outside our larger-scalable apertures, based on the tabulated
encircled energy corrections for point sources (Dressel et al.\ 2012).

For measurements of the flux in the ground-based observations or the
{\it Spitzer}/IRAC observations, we use the \textsc{mophongo} software
(Labb{\'e} et al.\ 2006, 2010a,b, 2013, 2015).  This software allows
to cope with the significant amounts of overlap in the light
distribution for nearby sources.  As with other software in the
literature with similar objectives, \textsc{mophongo} attempts to
overcome the issue of source confusion by assuming the high-resolution
{\it HST} images (here the $H_{160}$-band image) provide an accurate
model of spatial profile of sources in the ground-based/{\it
  Spitzer}/IRAC images and that only the normalization of source flux
varies from one passband to another.  \textsc{mophongo} then varies
their individual fluxes to obtain a good fit.  Measurements of the
flux for individual sources is then performed in fixed circular
apertures, after subtracting the model light profile from neighboring
sources.  $1.2''$-diameter, $1.8''$-diameter, and $2''$-diameter
apertures are used for the ground-based photometry, {\it
  Spitzer}/IRAC, and {\it Spitzer}/IRAC photometry over all of our
fields, the CANDELS GOODS-North+GOODS-South fields, and the CANDELS
UDS/COSMOS/EGS fields.  The measured fluxes are then corrected to
total based on the model profile for individual sources.  Narrower
apertures are used for our {\it Spitzer}/IRAC photometry over the
GOODS-North+GOODS-South fields to leverage the narrower FWHM of the
{\it Spitzer}/IRAC PSF in the Labb{\'e} et al.\ (2015) reductions.

\begin{figure*}
\epsscale{1.15} \plotone{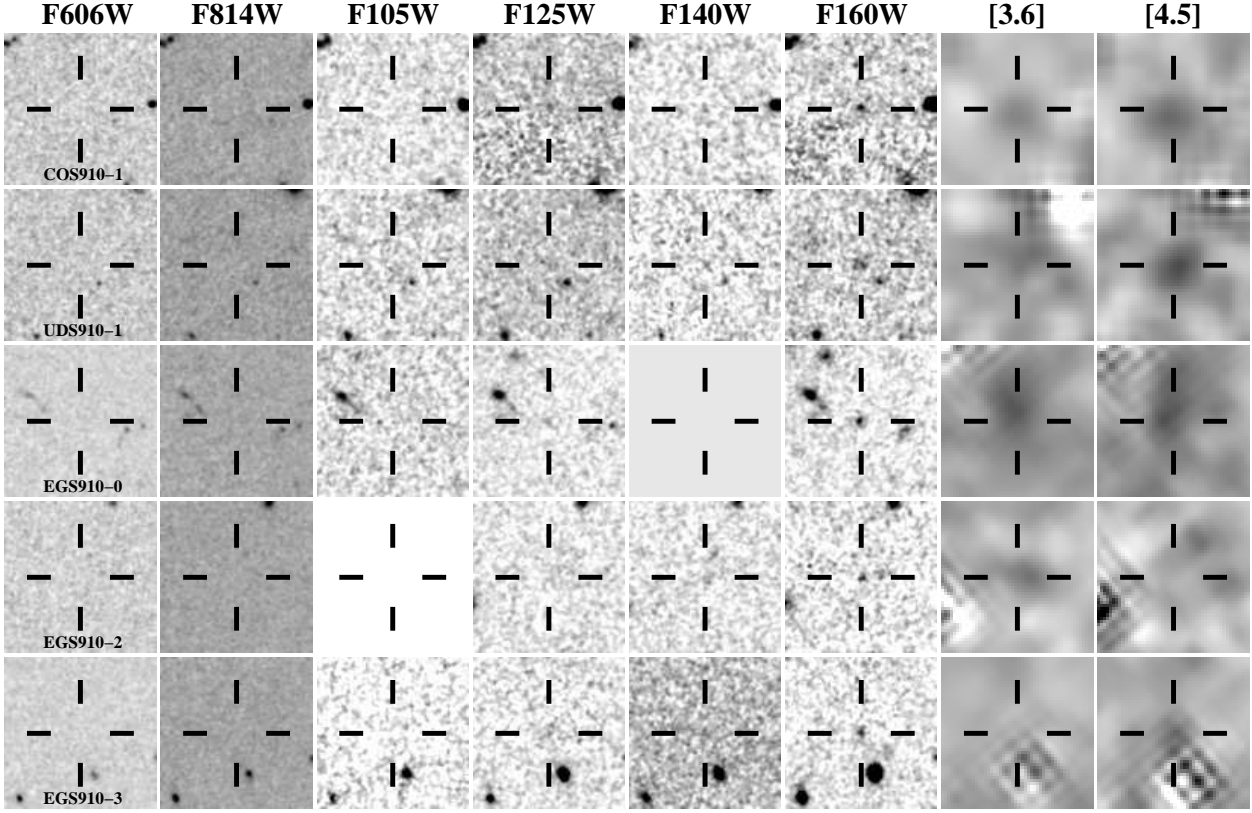}
\caption{{\it HST} + {\it Spitzer}/IRAC images for 5 candidate $z\sim9$-10
  galaxies which were confirmed as probable $z\geq9$ galaxies (or
  partially confirmed in the case of EGS910-2) using {\it HST} follow-up
  observations with our z9-CANDELS program.  Fits to the SEDs of these
  sources and the estimated redshift likelihood distributions are
  presented in Figure~\ref{fig:sed_conf}.\label{fig:stamp_conf}}
\end{figure*}

\begin{figure*}
\epsscale{1.18} \plotone{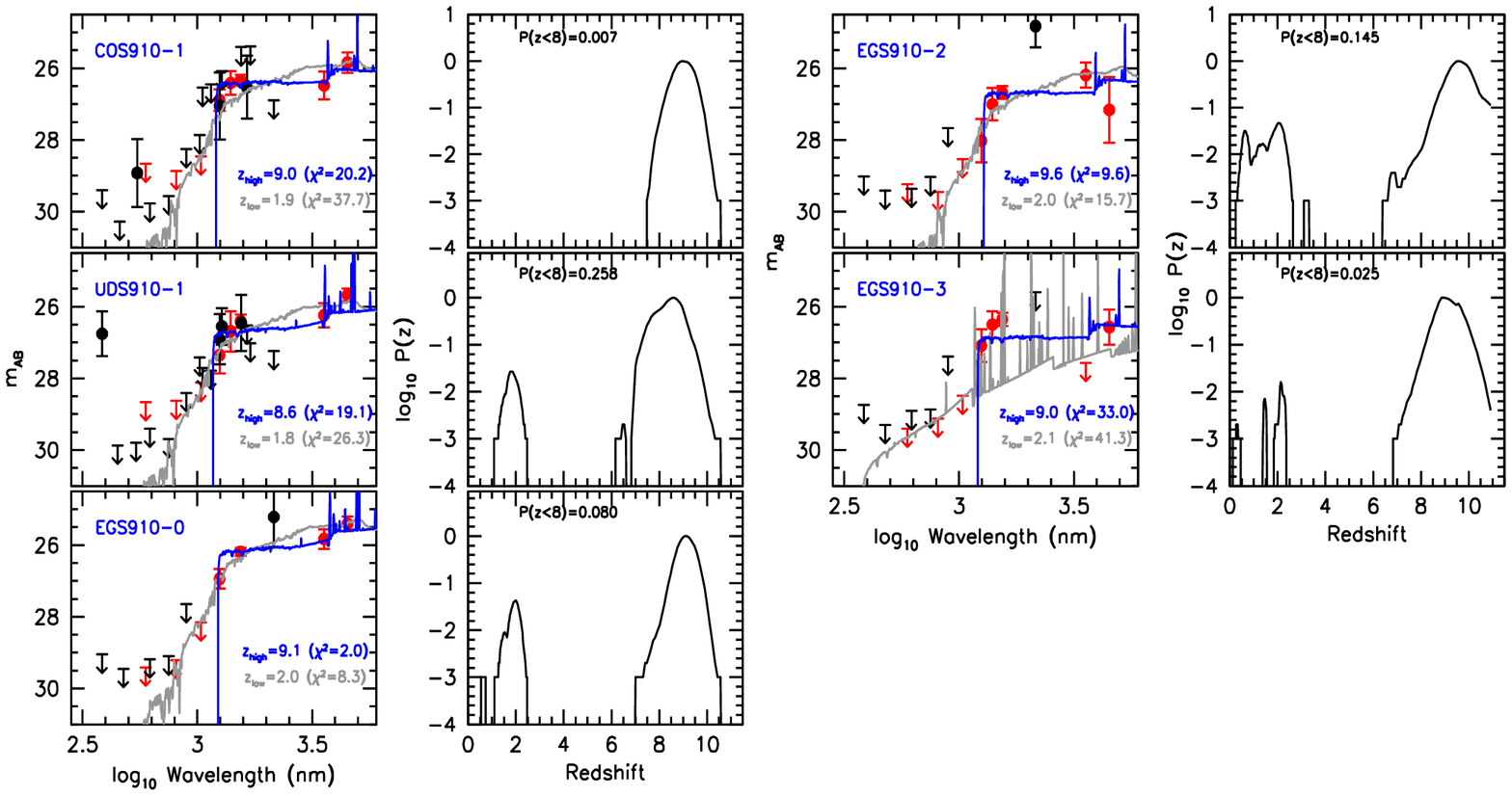}
\caption{(\textit{left}) Best-fit SED models to the observed {\it
    HST}+{\it Spitzer}/IRAC+ground-based photometry of five candidate
  $z\sim9$-10 galaxies (COS910-0, UDS910-1, EGS910-0, EGS910-2,
  EGS910-3) that have been photometrically confirmed (or partially
  confirmed in the case of UDS910-1 and EGS910-2) by observations from
  the z9-CANDELS follow-up program.  Red solid circles, $1\sigma$
  error bars, and $1\sigma$ limits are from the {\it HST} or {\it
    Spitzer}/IRAC observations, while the black solid circles,
  $1\sigma$ error bars, and $1\sigma$ limits are from the ground-based
  observations.  The solid blue line shows the best-fitting SED for a
  $z>6$ galaxy, while the grey line shows the best-fitting SED for a
  $z<6$ galaxy.  (\textit{right}) Redshift likelihood distribution for
  these $z\sim9$-10 candidates incorporating both our follow-up
  observations and the {\it HST}+{\it Spitzer}/IRAC+ground-based
  observations that were used in the pre-selection (\textit{solid
    lines}).\label{fig:sed_conf}}
\end{figure*}

\subsection{Selection of Bright $z\sim9$-10 Candidates over the CANDELS-UDS and 
CANDELS-COSMOS Fields}

\subsubsection{Selection Criteria} 

In searching for candidate galaxies at $z\sim9$-10, we suppose that
these galaxies have very similar colors and SEDs as galaxies at
slightly lower redshifts.  Specifically, we would expect these sources
to show a sharp spectral break at 1216\AA, due to strong absorption
from the neutral hydrogen forest, and to exhibit a blue $UV$-continuum
redward of the break.

For star-forming galaxies at redshifts $z\sim8.4$ and higher, the
Lyman-break will already have redshifted a significant way through the
$J_{125}$-band, yielding moderately red $J_{125}-H_{160}$ colors.  As
a result, the selection of all sources with red $J_{125}-H_{160}$
colors should allow us to identify the bulk of star-forming galaxies
from $z\sim8.7$ to $z\sim11$ (particularly if those galaxies are not
substantially dust obscured).

Here we search for candidate $z\gtrsim8.4$ galaxies over the
CANDELS-UDS and COSMOS fields using a $J_{125}-H_{160}>0.5$ criterion.
Star-forming galaxies with a $UV$-continuum slope $\beta$ of $-1.6$
(typical of luminous galaxies at $z=4$-7) would have $J_{125}-H_{160}$
color of $\sim0.5$ at $z=8.7$, but the lower-redshift limit for our
selection will depend on the intrinsic colors of individual galaxies
and also can be affected by observational noise.

In addition to our $J_{125}-H_{160}$ criterion, we also require that
sources be undetected ($<2\sigma$) in the $V_{606}$ or
$I_{814}$-bands.  Sources where the root mean square S/N in the
$V_{606}$ and $I_{814}$ bands is greater than 1 are excluded.  In
addition to these non-detection requirements on the {\it HST} optical
data, we also require that sources remain undetected ($<2.5\sigma$) in
an inverse-variance-weighted mean stack of the ground-based optical
data.

We also demand that sources show $H_{160}-[3.6]$ colors bluer than 1.4
mag to exclude intrinsically-red or old $z\sim2$ galaxies from our
samples, similar to the criteria that Oesch et al.\ (2014) or Bouwens
et al.\ (2015) apply.  This particular color cut corresponds to a
$UV$-continuum slope $\beta$ of 0.0 (where $f_{\lambda} \propto
\lambda^{\beta}$), which is approximately as red as bright galaxies
area observed to be at $z\sim6$-8 (e.g., Bouwens et al.\ 2012, 2014a;
Wilkins et al.\ 2011; Finkelstein et al.\ 2012; Rogers et al.\ 2014).

We require that all selected $z\sim9$-10 candidates show strong
evidence of corresponding to real sources.  We therefore require that
(1) sources be detected in the $H_{160}$ band at $>$5$\sigma$
significance, (2) the root mean square detection significance of
sources in the $JH_{140}$ and $H_{160}$-band images be at least 6, and
(3) the root mean square detection significance of sources in the
$JH_{140}$, [3.6], [4.5], and $K$-bands be at least $2\sigma$.

Finally, in the last step, we compute the redshift likelihood
distribution for each candidate source using the EAZY photometric
redshift code (Brammer et al.\ 2008) based on the photometry we have
available for sources, the standard EAZY\_v1.0 template set, and a
flat prior.  We supplemented the standard EAZY\_v1.0 template set with
SED templates from the Galaxy Evolutionary Synthesis Models (GALEV:
Kotulla et al.\ 2009).  Nebular continuum and emission lines were
added to the later templates using the Anders \& Fritze-v. Alvensleben
(2003) prescription, a $0.2 Z_{\odot}$ metallicity, and a rest-frame
EW for H$\alpha$ of 1300\AA$\,$(which appears to be appropriate for
$z\sim6$-7 (Smit et al.\ 2014, 2015; Roberts-Borsani et al.\ 2015).

The photometry utilized for constraining the likelihood distributions
for individual sources included the {\it HST}
$V_{606}I_{814}J_{125}JH_{140}H_{160}$ + Subaru-SuprimeCam $BgVriz$ +
CFHT/Megacam $ugriyz$ + UltraVISTA $YJHK_s$ + ZFOURGE $J_1 J_2 J_3 H_s
H_l$ + {\it Spitzer}/IRAC $3.6\mu$m+$4.5\mu$m S-CANDELS data sets for
the CANDELS COSMOS field, {\it HST}
$V_{606}I_{814}J_{125}JH_{140}H_{160}$ + Subaru-SuprimeCam $BVriz$ +
CFHT/Megacam $u$ + UKIRT/WFCAM $JHK_s$ + ZFOURGE $J_1 J_2 J_3 H_s H_l$
+ VLT/HAWKI/HUGS $YK_s$ data sets for the CANDELS UDS field, and the
{\it HST} $V_{606}I_{814}J_{125}JH_{140}H_{160}$ + CFHT/Megacam
$ugriyz$ + CFHT/WIRCam $K_s$ + {\it Spitzer}/IRAC 3.6$\mu$m+4.5$\mu$m data
sets for the CANDELS EGS field.  The depths of these observations are
provided in Table~\ref{tab:dataset} and their areal coverage is
illustrated in Figure~\ref{fig:layout}.

Sources that satisfied our aforementioned criteria, which showed a
$>$50\% probability of being at $z>8$, and which could be confirmed to
be a $>$90\% likelihood candidate with a single orbit of follow-up
observations (supposing sources are measured to have a flux of
0$\pm$12 nJy in the $Y_{105}$ band) made it into our final
preselection of candidate $z\sim9$-10 galaxies (to be targeted with
follow-up observations).  In computing the posterior probability that
a source has a redshift $z>8$ or $z<8$, we adopt a flat prior on the
redshift.

Table~\ref{tab:selcrit} provides a convenient compilation of all the
selection criteria we employed in preselecting candidate $z\sim9$-10
galaxies to follow-up with targeted observations.

\begin{deluxetable*}{ccccccc}
\tablewidth{0pt}
\tablecolumns{7}
\tabletypesize{\footnotesize}
\tablecaption{Photometrically-Confirmed $z\sim9$-10 Galaxies over the CANDELS Fields\label{tab:catalog_conf}}
\tablehead{
\colhead{ID} & \colhead{R.A.} & \colhead{Dec} & \colhead{$H_{160,AB}$} & \colhead{$z_{phot}$\tablenotemark{b}} & \colhead{P($z>8$)} & \colhead{Ref\tablenotemark{a}}}
\startdata
\multicolumn{7}{c}{$z\sim9$ Sample}\\
\multicolumn{3}{l}{Two-Part Search Strategy (Preselection + Follow-up: \S3, \S4):}\\
COS910-1 & 10:00:30.34 & 02:23:01.6 & 26.4$\pm$0.2 & 9.0$_{-0.5}^{+0.4}$ & 0.99 \\
EGS910-0 & 14:20:23.47 & 53:01:30.5 & 26.2$\pm$0.1 & 9.1$_{-0.4}^{+0.3}$ & 0.92 \\
EGS910-3 & 14:19:45.28 & 52:54:42.5 & 26.4$\pm$0.2 & 9.0$_{-0.7}^{+0.5}$ & 0.97 \\
UDS910-1\tablenotemark{c} & 02:17:21.96 & $-$05:08:14.7 & 26.6$\pm$0.2 & 8.6$_{-0.5}^{+0.6}$ & 0.74  \\\\
\multicolumn{3}{l}{Direct Search Strategy for $z\geq8.4$ Galaxies (\S5):}\\
GS-z9-1 & 03:32:32.05 & $-$27:50:41.7 & 26.6$\pm$0.2 & 9.3$\pm$0.5 & 0.9992 & [1], [2]\\
GS-z9-2  & 03:32:37.79 & $-$27:42:34.4 & 26.9 & 8.9$_{-0.3}^{+0.3}$ & 0.83 & [2]\\
GS-z9-3  & 03:32:34.99 & $-$27:49:21.6 & 26.9 & 8.8$_{-0.3}^{+0.3}$ & 0.95 & [2], [3]\\
GS-z9-4  & 03:33:07.58 & $-$27:50:55.0 & 26.8 & 8.4$_{-0.3}^{+0.2}$ & 0.97 & [2], [3]\\
GS-z9-5  & 03:32:39.96 & $-$27:42:01.9 & 26.4 & 8.7$_{-0.7}^{+0.8}$ & 0.55 & [2]\\
GN-z9-1 & 12:36:52.25 & 62:18:42.4 & 26.6$\pm$0.1 & 9.2$\pm$0.3 & $>$0.9999 & [1], [2]\\
\\
\multicolumn{7}{c}{$z\sim10$ Sample}\\
\multicolumn{3}{l}{Two-Part Search Strategy (Preselection + Follow-up: \S3, \S4):}\\
EGS910-2\tablenotemark{c} & 14:20:44.31 & 52:58:54.4 & 26.7$\pm$0.2 & 9.6$_{-0.5}^{+0.5}$ & 0.71 \\\\
\multicolumn{3}{l}{Direct Search Strategy for $z\geq8.4$ Galaxies (\S5):}\\
GN-z10-1\tablenotemark{d} & 12:36:25.46 & 62:14:31.4 & 26.0$\pm$0.1 & 11.1$\pm$0.1 & $>$0.9999 & [1], [2], [4], [5]\\
GN-z10-2 & 12:37:22.74 & 62:14:22.4 & 26.8$\pm$0.1 & 9.9$\pm$0.3 & 0.9994 & [1], [2]\\
GN-z10-3 & 12:36:04.09 & 62:14:29.6 & 26.8$\pm$0.2 & 9.5$\pm$0.4 & 0.9981 & [1], [2]\\
GS-z10-1 & 03:32:26.97 & $-$27:46:28.3 & 26.9$\pm$0.2 & 9.9$\pm$0.5 & 0.9988 & [1], [2]
\enddata
\tablenotetext{a}{References: [1] Oesch et al.\ 2014, [2] Bouwens et al.\ 2015, [3] McLure et al.\ 2013, [4] Oesch et al.\ 2016, [5] Bouwens et al.\ 2010}
\tablenotetext{b}{$1\sigma$ uncertainties are computed based on the $z>4$ likelihood distributions.} 
\tablenotetext{c}{This candidate could only be partially confirmed,
  given the limited orbit allocation to our {\it HST} program.}

\tablenotetext{d}{This source is now spectroscopically confirmed to
  lie at $z=11.1$ (Oesch et al.\ 2016), but broadly lies within our
  $z\sim10$ selection window.}

\end{deluxetable*}

\subsubsection{UDS+COSMOS Results}

Applying the selection criteria from the previous section to our
source catalogs over the CANDELS-UDS and CANDELS-COSMOS fields, we
found five sources which satisfied all of the criteria.  A list of all
6 sources satisfying these criteria are included in
Table~\ref{tab:catalog} and similar candidates from the CANDELS-EGS
field.

The observed spectral energy distributions for these five candidate
$z\sim9$-10 galaxies are presented in Appendix A
(Figure~\ref{fig:sed1}), along with SED fits to a model $z>8$ galaxy
and a model $z<3$ galaxy.  Also shown on this figure is the redshift
likelihood distribution (\textit{solid black line}) based on the
photometry we have available for each candidate in the $\sim$20
different wavelength channels ({\it HST} + {\it Spitzer}/IRAC + ground-based
observations).  In addition, this figure presents the redshift
likelihood distribution we would expect, assuming these candidates are
not detected in the single orbit of follow-up $Y_{105}$-band
observations from the z9-CANDELS program.

Postage stamp images of these six candidates are also presented in
Appendix A (Figure~\ref{fig:stamp1}).  As should be obvious from this
figure, all six of the present $z\sim9$-10 candidates show clear
detections in the $H_{160}$-band, as well as significant
$\sim2$-3$\sigma$ detections in $J_{125}$-band and $JH_{140}$-band
observations (where available), as well as in the S-CANDELS
{\it Spitzer}/IRAC data.

All six of these candidates bear a remarkable similarity to the first
samples of particularly luminous $z\sim9$-10 galaxies identified by
Oesch et al.\ (2014) in terms of their very blue $H_{160}-[3.6]$
colors (see also Wilkins et al.\ 2016), red $[3.6]-[4.5]$ colors, and
observed sizes (Holwerda et al.\ 2015).

\subsection{Selection of $z\sim9$-10 Candidates over the CANDELS-EGS}

\subsubsection{Selection Criteria}

The selection of candidate $z\sim9$-10 galaxies over the CANDELS-EGS
field is even more challenging than over the CANDELS-UDS and
CANDELS-COSMOS fields due to the lack of deep observations at
1.05$\mu$m over the CANDELS-EGS field.  $Y$-band observations (at
$1.05\mu$m) play a crucial role in excluding the possibility that
sources can correspond to slightly reddened star-forming galaxies at
$z\sim7.5$-8.5 or correspond to passive or reddened galaxies at much
lower redshifts.

In selecting $z\sim9$-10 candidates over the CANDELS-EGS field, we
therefore adopted almost identical criteria as over the CANDELS-COSMOS
or CANDELS-UDS fields, with one exception.  Instead of requiring
sources to have a $>$50\% probability of corresponding to a $z>8$
galaxy, we required that sources be capable of confirmation with a
single orbit of {\it HST} observations at $1.05\mu$m.  For the purposes of
selection, we take confirmation to correspond to the source having
$>$90\% likelihood of being at $z>8$ after adding a flux constraint of
0$\pm$12 nJy to the observed SED at $1.05\mu$m (though we obtained
follow-up observations in $JH_{140}$ for the one case where the
$J_{125}-H_{160}$ color was $>$1.2).

\subsubsection{EGS Results}

Applying the selection criteria from the previous section to our
source catalogs over the CANDELS-EGS field, we found six additional
sources which satisfied all of the criteria (Table~\ref{tab:catalog}).

The observed spectral energy distributions for these six candidate
$z\sim9$-10 galaxies are presented in Appendix A
(Figure~\ref{fig:sed1}), along with SED fits to a model $z>6$ galaxy
and a model $z<6$ galaxy.  Postage stamp images of these six candidate
$z\sim9$-10 galaxies are also provided.

The most promising $z\sim9$-10 candidates we identified over the
CANDELS-EGS field were the EGS910-0, EGS910-2, and EGS910-3.  All 3
sources show evidence for a sharp break break at $1.2\mu$m, as well as
a blue spectral slope redward of the break.  The other candidates also
show evidence for a strong spectral break at 1.2$\mu$m and a blue
spectral slope redward of the break, but also show possible flux in
$\sim$1-2 passbands blueward of the break.  Until observations from
our z9-CANDELS follow-up program became available on these
candidates, it was not possible to determine whether they are more
likely to correspond to bona-fide $z\sim9$-10 galaxies or $z\sim1$-3
interlopers.

\begin{figure*}
\epsscale{1.18} \plotone{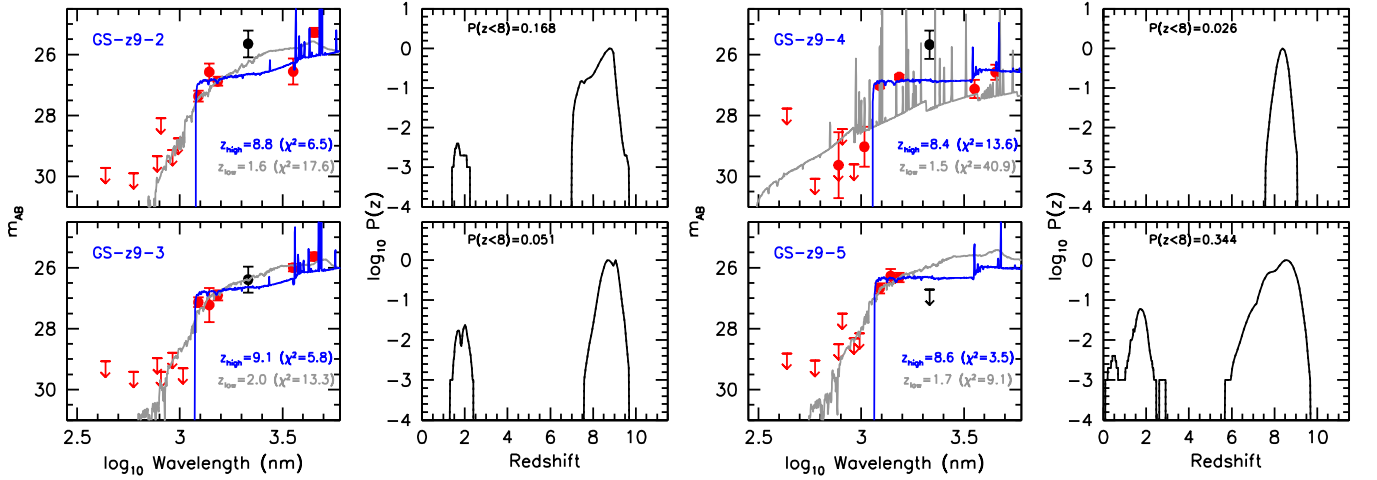}
\caption{(\textit{left}) Best-fit SED models to the observed {\it
    HST}+{\it Spitzer}/IRAC+ground-based photometry of three candidate
  $z\sim9$ galaxies (GS-z9-2, GS-z9-3, GS-z9-4) that satisfied our
  criteria for selection.  This figure also includes another candidate
  $z\sim9$ galaxy (GS-z9-5), but appears to nevertheless be a
  high-redshift galaxy.  {\it Spitzer}/IRAC observations are not shown
  for GS-z9-5 since it is nearby a very bright star and under a
  diffraction spike.  These sources were identified in a separate
  search over the extended GOODS-South area (ERS, CANDELS GOODS-South,
  HUDF09-1, HUDF09-2: see \S5.1).  The points and lines are otherwise
  as in Figure~\ref{fig:sed_conf}.  (\textit{right}) Redshift
  likelihood distribution for these $z\sim9$ candidates (\textit{solid
    lines}).\label{fig:sed_south}}
\end{figure*}

\begin{figure*}
\epsscale{1.15} \plotone{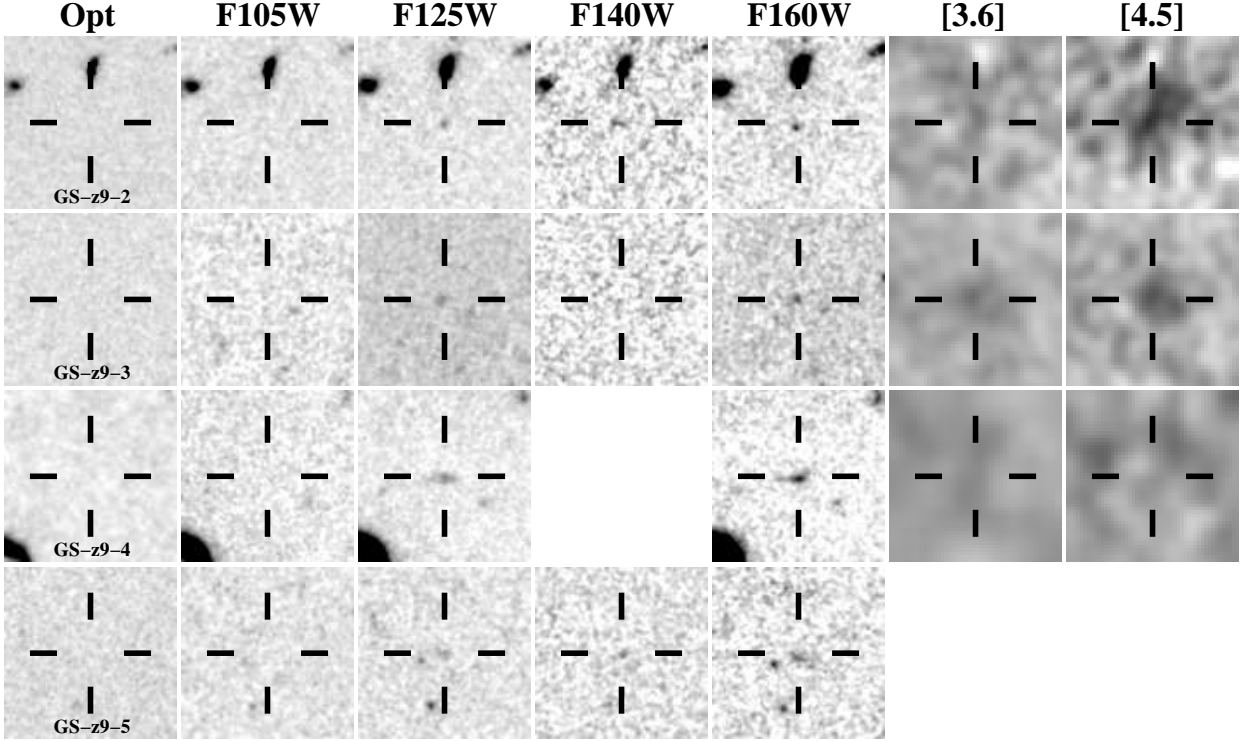}
\caption{{\it HST} + {\it Spitzer}/IRAC images ($6''\times6''$) of
  three candidate $z\sim9$ galaxies (GS-z9-2, GS-z9-3, GS-z9-4) that
  satisfied our criteria for selection and another candidate $z\sim9$
  galaxy (GS-z9-5) that did not satisfy these criteria (but appears
  nevertheless to be a high-redshift source).  {\it Spitzer}/IRAC
  observations are not shown for GS-z9-5 since it is nearby a very
  bright star and under a diffraction spike.  We have identified these
  candidates in a separate search over the extended GOODS-South area
  (ERS, CANDELS GOODS-South, HUDF09-1, HUDF09-2: see
  \S5.1).\label{fig:stamp_south}}
\end{figure*}

\section{Nature of the Targeted $z\sim9$-10 Candidates}

{\it HST} observations are now available over all 12 candidate
$z\sim9$-10 galaxies targeted by our z9-CANDELS program.  These
observations allow us to make a fairly definitive assessment of the
nature of these candidate $z\sim9$-10 based on the flux we measure for
these candidates at $1.05\mu$m.  One orbit of $Y_{105}$-band
observations have already been obtained for eight candidates targeted
by our program COS910-0, COS910-1, COS910-3, UDS910-0, UDS910-1,
EGS910-0, EGS910-1, EGS910-3, and EGS910-4.  Slightly shallower
observations (i.e., 1/3 and 2/3 of an orbit) in the $Y_{105}$-band
were acquired on the candidates EGS910-5 and COS910-2, due to the
greater brightness of the former candidate and the utility of an
additional 1/3 orbit $JH_{140}$-band observations to investigate the
nature of the potential $z\sim10$ candidate galaxies EGS910-2 and
COS910-3.

$Y_{105}$-band images for these candidates are presented in either
Figures~\ref{fig:stamp_conf} or \ref{fig:stamp_notconf} from Appendix
B, in conjunction with images of these candidates at other
wavelengths.  Figures~\ref{fig:sed_conf} and \ref{fig:sed_notconf}
from Appendix B show the observed SEDs for the targetted $z\sim9$-10
candidates in CANDELS program.

The present observations confirm photometrically 5 of the first 12
$z\sim9$-10 candidates targeted by our program.  Two of these five
confirmations are only partial confirmations (EGS910-2 and UDS910-1:
as more observations are needed for these candidates to $>$90\% secure).
Detailed remarks on the confirmed $z\sim9$-10 candidates can be found
here:\\
\noindent \textbf{COS910-1:} COS910-1 is not detected ($<1\sigma$) in
the $Y_{105}$-band follow-up observations at $1.05\mu$m.  A detailed
fit to its SED suggests that it is actually a star-forming galaxy at
$z=9.1$, with $<$0.7\% probability of it corresponding to a $z<8$
galaxy.\\
\noindent \textbf{UDS910-1:} UDS910-1 is not detected at
($<$1$\sigma$) in the $Y_{105}$-band follow-up observations we
obtained at 1.05$\mu$m.  Rederiving the redshift likelihood
distribution using the new flux information in the $Y_{105}$-band, we
compute a best-fit photometric redshift of 8.6, with a 4\% and 24\%
probability of corresponding to a $z<7$ and $z<8$ source,
respectively.\\
\noindent \textbf{EGS910-0:} EGS910-0 is not detected ($<$1$\sigma$)
in the $Y_{105}$-band follow-up observations at $1.05\mu$m.
Rederiving the redshift likelihood distribution using the new flux
information in the $Y_{105}$-band, we compute a best-fit photometric
redshift of 9.1, with only a 4\% probability of corresponding to a
$z<8$ source.\\
\noindent \textbf{EGS910-2:} Follow-up of EGS910-2 in the
$JH_{140}$-band shows a clear $2.6\sigma$ detection of the source and
which is in excellent agreement with the expected flux given a model
redshift of $z\sim9.6$ for the source.  Nevertheless, the source is
sufficiently faint that the redshift likelihood distribution shows a
29\% likelihood of the source being at $z<8$.  Deeper follow-up
observations at $1.05\mu$m will be required to rule out the $z<8$
solution.\\
\noindent \textbf{EGS910-3:} EGS910-3 shows no detection
($<$1$\sigma$) in the $Y_{105}$-band follow-up observations we
obtained at 1.05$\mu$m.  Rederiving the redshift likelihood
distribution using the new flux information in the $Y_{105}$-band, we
compute a best-fit photometric redshift of 9.0, with a 3\% probability
of corresponding to a $z<8$ source.

The 3 $z\gtrsim8.4$ candidates confirmed -- and 2 candidates partially
confirmed -- by our z9-CANDELS program are compiled for convenience in
Table~\ref{tab:catalog_conf}.  We will also include in this table some
additional candidates we identify in \S5 (and also as identified by
Oesch et al.\ 2014 and Bouwens et al.\ 2015).

Our overall confirmation rate is 42\% (5/12) for sources preselected
by our criteria.  We achieve an even higher 56\% success rate
targeting those sources from our selection which are high-probability
($>$50\%) $z>8$ galaxies, before our follow-up observations.  While
imperfect, this program is very efficient, supplementing some 270
orbits of {\it HST} time and hundreds of hours of {\it Spitzer} time
with only 11 orbits of additional {\it HST} time.  By contrast, the
CANDELS + ERS programs over the GOODS-North + South cost some
$\sim$500 orbits, and we identified only 9 candidates in those data,
or 0.02 $z\sim9$-10 candidate per invested orbit.

Detailed remarks on the $z\sim9$-10 candidate galaxies which were not
confirmed by our follow-up program are included in Appendix A.

\section{Completing the Census of Candidate $z\sim9$ Galaxies over the CANDELS GOODS-North, GOODS-South, and ERS Fields}

We can obtain the best constraints on the volume density of bright
$z\sim9$-10 galaxy candidates by not simply considering a search over
the CANDELS UDS, COSMOS, and EGS fields as we did in the previous
sections, but also considering a search for similar sources over the
GOODS-North and GOODS-South fields.

\subsection{Criteria for Identifying $z\sim8.5$-9.0 Galaxies}

The purpose of the present section is to obtain a complete census of
the bright $z\sim8.5$-11 galaxy candidates over the CANDELS
GOODS-North+GOODS-South + ERS fields.

In Oesch et al.\ (2014) and Bouwens et al.\ (2015), we already
conducted a significant search for galaxies in this redshift range, by
looking for sources with red $J_{125}-H_{160}>0.5$ colors and blue
$H_{160}-[3.6]<1.4$ colors.  However, such a selection is only
sensitive to galaxies with redshifts $z\gtrsim9$ and can suffer
significant incompleteness at $z<9$.

Here we extend the search from Oesch et al.\ (2014) and Bouwens et
al.\ (2015) to also consider sources with redshifts $z\gtrsim 8.4$.
We select these sources by considering all those sources which satisfy
the $z\sim8$ color-color criteria of Bouwens et al.\ (2015), deriving
photometric redshifts for all such sources using the EAZY photometric
redshift code (Brammer et al.\ 2008), and including those sources
where the most likely redshift is greater than 8.4.

The photometry we consider in deriving the redshift likelihood
contours are the Bouwens et al.\ (2015) reductions of the {\it HST}
$B_{435}V_{606}i_{775}I_{814}z_{850}Y_{098}Y_{105}J_{125}JH_{140}H_{160}$
data, the Labb{\'e} et al.\ (2015) reductions of essentially all
{\it Spitzer}/IRAC observations over the GOODS-North and South fields, and
the Bouwens et al.\ (2015) reductions of the HUGS HAWK-I $K_{s}$-band
observations.

Briefly, the Bouwens et al.\ (2015) selection criteria for identifying
$z\sim8$ sources is 
\begin{eqnarray*}
(Y_{105}-J_{125}>0.45)\wedge (J_{125}-H_{160}<0.5) \wedge \\
(Y_{105}-J_{125} > 0.75(J_{125}-H_{160})+0.525)
\end{eqnarray*} 
for sources over the CANDELS GOODS-North + GOODS-South fields and
\begin{eqnarray*}
(Y_{098}-J_{125}>1.3)\wedge (J_{125}-H_{160}<0.5) \wedge \\
(Y_{098}-J_{125} > 0.75(J_{125}-H_{160})+1.3)
\end{eqnarray*}
for sources over the $\sim$40 arcmin$^2$ ERS field.  Sources are
required to be detected at $6\sigma$ in a $\chi^2$ stack of the
$H_{160}$-band or $JH_{140}+H_{160}$ band observations redward of the
break (in a fixed 0.36$''$-diameter aperture).

To ensure contamination is kept to a minimum, an optical ``$\chi^2$''
is computed for each candidate source (Bouwens et al.\ 2011b) based on
the flux in the $B_{435}V_{606}i_{775}I_{814}z_{850}$-band
observations.  $\chi_{opt} ^2$ is taken to equal $\Sigma_{i}
\textrm{SGN}(f_{i}) (f_{i}/\sigma_{i})^2$ where $f_{i}$ is the flux in
band $i$ in a consistent aperture, $\sigma_i$ is the uncertainty in
this flux, and SGN($f_{i}$) is equal to 1 if $f_{i}>0$ and $-1$ if
$f_{i}<0$.  Any candidate with a measured $\chi_{opt}$ in excess of 4
is excluded from our selections.

\begin{figure}
\epsscale{1.18} \plotone{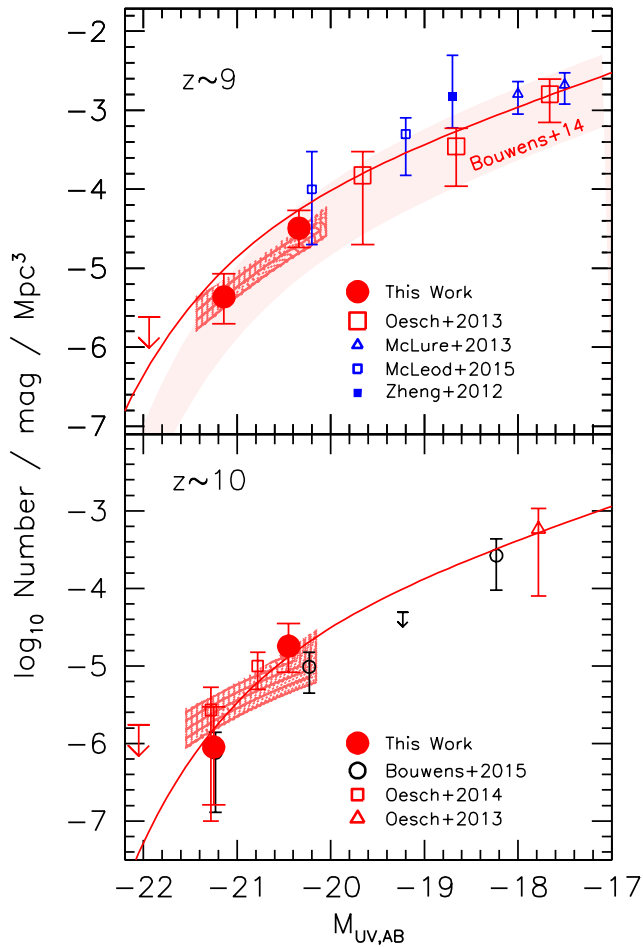}
\caption{Simple binned determinations of the $UV$ LF for luminous
  galaxies at $z=9$ (\textit{upper panel}) and $z=10$ (\textit{lower
    panel}).  $1\sigma$ upper limits on the volume density of $z\sim9$
  and $z\sim10$ galaxies are included at $\sim$$-22$ mag.  The shaded
  hatched red region indicates the volume densities (at a given
  $M_{UV}$) preferred at 68\% confidence by this analysis (see
  Table~\ref{tab:modellf}).  To put these constraints on the bright
  end of the $UV$ LF in context, we also include determinations of the
  $z\sim9$ and $z\sim10$ $UV$ LFs at lower luminosities from Zheng et
  al.\ (2012: \textit{solid blue square}), Oesch et al.\ (2013:
  \textit{open red square}), McLure et al.\ (2013: \textit{open blue
    triangles}), and McLeod et al.\ (2015: \textit{open blue
    squares}).  The lightly-shaded red region shows the constraints on
  the $z\sim9$ LF as derived by Bouwens et al.\ (2014b) using a search
  for $z\sim9$ galaxies over the CLASH program (Postman et al.\ 2012).
  The overplotted line shows an extrapolation of the Bouwens et
  al.\ (2015) LF results to $z\sim9$ and $z\sim10.2$ based on the
  fitting formula provided in \S7.1.\label{fig:complf}}
\end{figure}

We only search for $z\gtrsim8.4$ sources over the GOODS-North and
GOODS-South fields brightward of $H_{160,AB}=27$ mag to ensure that we
have strong constraints on the nature of the selected sources to the
limit of our search.  The effective depth of our $z\sim9$-10 search
over the CANDELS-UDS, COSMOS, and EGS fields is also approximately
$\sim$27 mag, so the effective depth of our search is similar across
all five CANDELS fields that we utilize.

\subsection{Selection Results}

Using the selection criteria from the previous section, we identify
three high-probability ($>$90\% confidence) and one moderate
probability ($\sim$50\% confidence) $z\sim8.4$-9.0 galaxies over the
ERS, CANDELS GOODS-South, and CANDELS GOODS-South fields.

The $H_{160}$-band magnitudes of the $z\sim8.4$-9.0 galaxies we have
selected range from 26.4 and 26.9, similar to that found for our
$z\sim9$-10 sample over the CANDELS-UDS, COSMOS, and EGS fields.  We
have included the four new $z\sim8.4$-9.0 candidates in
Table~\ref{tab:catalog_conf}, along with other high-probability
$z\sim8.4$-11 identified here.  Fits to the observed SEDs for our new
$z\sim8.4$-9.0 candidates over these fields are shown in
Figure~\ref{fig:sed_south}.  Postage stamp images of the candidates
are provided in Figure~\ref{fig:stamp_south}.

\subsection{Criteria for Identifying $z\gtrsim9.0$ Galaxies}

As performed by Oesch et al.\ (2014) and Bouwens et al.\ (2015), we
also include sources with $J_{125}-H_{160}>0.5$, $H_{160}-[3.6]<1.4$
colors.  Our selection criteria for identifying these sources are
essentially identical to that utilized by Bouwens et al.\ (2015: see
also Oesch et al.\ 2014) except that a $J_{125}-H_{160}>0.5$ color
criterion is utilized.

We identify exactly the same set of sources as Oesch et al.\ (2014)
identify using the above criteria.  A compilation of these sources and
other $z\sim8.4$-9.0 sources identified over the CANDELS GOODS-North,
GOODS-South, and ERS fields is provided in
Table~\ref{tab:catalog_conf}.

\section{Impact of Gravitational Lensing from Foreground Galaxies on These Results}

From previous work (Wyithe et al.\ 2011; Barone-Nugent et al.\ 2015;
Mason et al.\ 2015; Fialkov \& Loeb 2015), it is well known that
gravitational lensing from foreground galaxies can have a particularly
significant effect in enhancing the surface density of bright $z\geq
6$ galaxies on the sky.  This is especially true for the brightest
sources due to the intrinsic rarity and the large path length
available for lensing by foreground sources.

Given this phenomenon, it has become increasingly common for
researchers searching for the brightest $z\sim6$-10 galaxies to look
for possible evidence of lensing amplification (Oesch et al.\ 2014;
Bowler et al.\ 2014, 2015; Zitrin et al.\ 2015; Roberts-Borsani et
al.\ 2016).  While there are a number of cases where such
magnification boosts may be present (e.g., Barone-Nugent et al.\ 2015;
Roberts-Borsani et al. 2016), the fraction of lensed sources among
bright samples still does not appear to be particularly high (Bowler
et al.\ 2015).

As in the above work, we explicitly check our compilation of bright
$z\sim9$-10 galaxy candidates from these fields for evidence of
gravitational lensing.  For convenience, we use the Skelton et
al.\ (2014) catalogs providing radii and stellar mass estimates for
all sources over the CANDELS areas we have searched.  The Skelton et
al.\ (2014) catalogs use the diverse multi-wavelength data over the
CANDELS fields, including {\it HST} optical, near-infrared, {\it
  Spitzer}/IRAC, and ground-based observations, to provide flux
measurements of a wide wavelength range and then use these flux
measurements to estimate the redshifts and stellar masses.

As in Roberts-Borsani et al.\ (2016), we model galaxies in our bright
$z\sim9$-10 sample as singular isothermal spheres, and we use the
measured half-light radius and inferred stellar mass to derive a
velocity dispersion estimates for individual galaxies in these
samples.  We found only two examples of galaxies whose measured
fluxes appear likely to be slightly boosted ($>$0.1 mag) by lensing
amplification:\\

\noindent \textbf{EGS910-3:} There is a foreground galaxy at
$z\sim1.9$ with estimated stellar mass of a 10$^{10.32}$-$M_{\odot}$
that lies within 1.9 arcsec of this source.  Based on the velocity
dispersion we estimate for this source of $\sim$220 km/s, we compute a
magnification boost of 0.25 mag for this source.

\noindent \textbf{GN-z10-2:} This source is estimated to be boosted by
0.11 mag by a $10^{10.64}$ $M_{\odot}$ galaxy with a spectroscopic
redshift of $z=1.02$ (Barger et al.\ 2008) that lies within 4.0$''$ of
the targeted source.  This source was previously flagged as being
slightly lensed by Oesch et al.\ (2014).

\section{Implications of Our Search Results}

\subsection{Constraints on the $UV$ Luminosity Functions at $z=9$ and $z=10$}

In this section, we utilize the combined sample of $z\sim9$-10
candidates over the CANDELS-UDS, CANDELS-COSMOS, and CANDELS-EGS
fields and similar $z\sim9$-10 candidates over CANDELS GOODS-North and
GOODS-South fields (\S5) to quantify the $UV$ LFs at $z\sim9$ and
$z\sim10$.  Table~\ref{tab:catalog_conf} provides a compilation of the
relevant sources for our determination of the LF.

\begin{figure}
\epsscale{1.18} \plotone{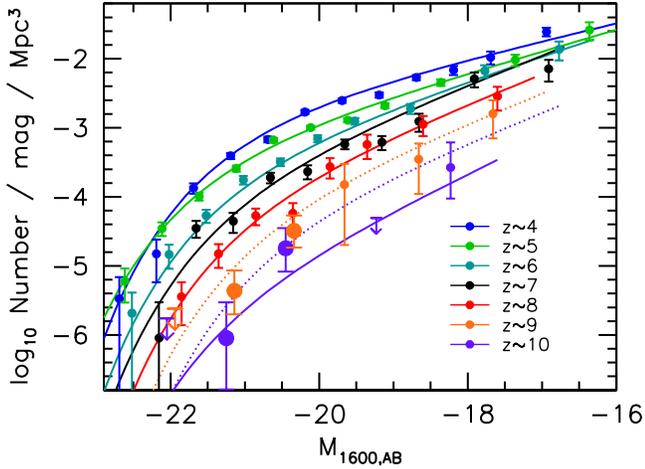}
\caption{Present determinations of the bright end of the $UV$ LF at
  $z\sim9$ and $z\sim10$ using all five CANDELS fields (\textit{orange
    and dark purple circles, respectively, with $1\sigma$
    uncertainties}).  $1\sigma$ upper limits are as in
  Figure~\ref{fig:complf}.  For context, we also include fainter
  determinations of the $UV$ LFs at $z\sim9$ and $z\sim10$ from Oesch
  et al.\ (2013) and Bouwens et al.\ (2015: \textit{small circles} and
  \text{solid line}).  The dashed lines indicate extrapolations of the
  Bouwens et al.\ (2015) LF relations to $z\sim9$ and $z\sim10.2$ (see
  \S7.1).  The $z\sim4$, $z\sim5$, $z\sim6$, $z\sim7$, and $z\sim8$
  LF determinations from Bouwens et al.\ (2015) are also shown.\label{fig:lfall}}
\end{figure}

\begin{deluxetable}{lc}
\tablewidth{0pt}
\tabletypesize{\footnotesize}
\tablecaption{Binned Determination of the rest-frame $UV$ LF at $z\sim9$ and $z\sim10$.\label{tab:swlf}}
\tablehead{
\colhead{$M_{1600,AB}$\tablenotemark{a}} & \colhead{$\phi_k$ (10$^{-3}$ Mpc$^{-3}$ mag$^{-1}$)}}
\startdata
\multicolumn{2}{c}{$z\sim9$ galaxies}\\
$-$21.94 & $<$0.0024\tablenotemark{b}\\
$-$21.14 & 0.0044$_{-0.0024}^{+0.0042}$\\
$-$20.34 & 0.0322$_{-0.0138}^{+0.0217}$\\
\multicolumn{2}{c}{$z\sim10$ galaxies}\\
$-$22.05 & $<$0.0017\tablenotemark{b}\\
$-$21.25 & 0.0009$_{-0.0007}^{+0.0021}$\\
$-$20.45 & 0.0180$_{-0.0098}^{+0.0174}$
\enddata 
\tablenotetext{a}{Derived at a rest-frame wavelength of 1600\AA.}
\tablenotetext{b}{$1\sigma$ upper limit.}
\end{deluxetable}

As in our recent paper on the $z\sim4$-10 LFs, we use the results from
these simulations to derive the selection volumes needed to relate the
$UV$ LF function $\phi(M)$ to the observed surface density of sources
on the sky.  Formally, we write the $UV$ LF in stepwise format
$\phi_j$ as $\Sigma \phi_j W(M-M_j)$ where $j$ is an index running
over the magnitude bins, where $M_j$ corresponds to the absolute
magnitude at the center of each bin, where
\begin{equation}
W(x) = 
\begin{array}{cc} 
0, & x < -0.4\\
1, & -0.4 < x < 0.4\\
0, & x > 0.4.
\end{array}
\end{equation}
and where $x$ gives the position within a magnitude bin.  We take the
width of magnitude bins to be 0.8 mag (e.g. versus the 0.5-mag used by
Bouwens et al.\ 2015), given the limited number of bright $z\sim9$ and
$z\sim10$ galaxies.

We then look for the derived LF $\phi_j$ that yields the observed
surface density of $z\sim9$ and $z\sim10$ galaxies on the sky with
maximum probability $\cal L$.  As in Bouwens et al.\ (2015), the
likelihood $\cal L$ is computed as
\begin{equation}
{\cal L}=\Pi_{field} \Pi_i p(m_i)
\label{eq:ml}
\end{equation}
where the above products runs over the different search fields and
magnitude interval $i$ used in the LF determinations, and where
$p(m_i)$ is probability of identifying a certain number of sources in
magnitude interval $i$ in a given search field.

For simplicity (and given the small numbers in each of our samples:
see the discussion in \S4 of Bouwens et al.\ 2008), we ignore
field-to-field variance in deriving the LF results and compute the
likelihood that our survey fields show a certain number of sources
assuming Poissonian statistics.  We therefore compute $p(m_i)$ as
follows:
\begin{equation}
p(m_i) = e^{-N_{exp,i}} \frac{(N_{exp,i})^{N_{obs,i,j}}}{(N_{obs,i,j})!}
\label{eq:pmi}
\end{equation}
where $N_{obs,i}$ is the observed number of sources in search field
and magnitude interval $i$, where $N_{exp,i}$ is the expected number
of sources in a search field and magnitude interval $i$.  The expected
number of sources in a search field $N_{expected,i}$ is computed as
\begin{equation}
N_{expected,i} = \Sigma _{j} \phi_j V_{i,j}
\label{eq:numcountg}
\end{equation}
where $V_{i,j}$ is the effective volume over which one could expect to
find a source of absolute magnitude $j$ in the observed magnitude
interval $i$.  

\begin{deluxetable}{lccccc}
\tablewidth{0pt}
\tabletypesize{\footnotesize}
\tablecaption{68\% confidence regions on the volume density of galaxies at $z\sim9$ and $z\sim 10$ vs. $M_{UV}$\label{tab:modellf}}
\tablehead{
\colhead{} & \multicolumn{2}{c}{Volume Density (10$^{-3}$ Mpc$^{-3}$ mag$^{-1}$)} \\
\colhead{$M_{1600,AB}$\tablenotemark{a}} & \colhead{Lower Bound\tablenotemark{b}} & \colhead{Upper Bound\tablenotemark{b}}}
\startdata
\multicolumn{3}{c}{$z\sim9$ galaxies}\\
$-$21.84 & 0.0006 & 0.0024 \\
$-$21.64 & 0.0010 & 0.0032 \\
$-$21.44 & 0.0016 & 0.0044 \\
$-$21.24 & 0.0024 & 0.0060 \\
$-$21.04 & 0.0039 & 0.0084 \\
$-$20.84 & 0.0059 & 0.0120 \\
$-$20.64 & 0.0087 & 0.0174 \\
$-$20.44 & 0.0126 & 0.0262 \\
$-$20.24 & 0.0176 & 0.0402 \\
$-$20.04 & 0.0250 & 0.0621 \\\\
\multicolumn{3}{c}{$z\sim10$ galaxies}\\
$-$21.95 & 0.0004 & 0.0018 \\
$-$21.75 & 0.0006 & 0.0022 \\
$-$21.55 & 0.0009 & 0.0028 \\
$-$21.35 & 0.0012 & 0.0036 \\
$-$21.15 & 0.0017 & 0.0048 \\
$-$20.95 & 0.0023 & 0.0066 \\
$-$20.75 & 0.0030 & 0.0093 \\
$-$20.55 & 0.0039 & 0.0135 \\
$-$20.35 & 0.0049 & 0.0200 \\
$-$20.15 & 0.0060 & 0.0299 
\enddata 
\tablenotetext{a}{Derived at a rest-frame wavelength of 1600\AA.}
\tablenotetext{b}{68\% Confidence Region}
\end{deluxetable}

The selection volumes $V_{i,j}$ are estimated using almost an
identical procedure to that in Bouwens et al.\ (2015).  Specifically,
we constructed catalogs with mock sources spanning the entire redshift
range $z\sim7.5$ to $z\sim12$.  To ensure that sources had reasonable
sizes and morphologies, we randomly selected similar luminosity
$z\sim4$ galaxies from the Hubble Ultra Deep Field (Beckwith et
al.\ 2006; Illingworth et al.\ 2013) to use as a template to modeling
the two-dimensional pixel-by-pixel profiles of individual sources.
The size of the model sources were assumed to scale with redshift as
$(1+z)^{-1.2}$ to match the size scaling observed for sources with
fixed luminosity from $z\sim10$ to $z\sim2$ (Bouwens et al.\ 2015;
Oesch et al.\ 2010; Ono et al.\ 2013; Holwerda et al.\ 2014; Kawamata
et al.\ 2015; Shibuya et al.\ 2015).  $UV$ continuum slopes of sources
were assumed to have a mean value of $-1.8$, consistent with that
measured at high luminosities at $z\sim5$-8 (Bouwens et al.\ 2012,
2014; Finkelstein et al.\ 2012; Willott et al.\ 2013; Rogers et
al.\ 2014), with a dispersion of 0.3 (Bouwens et al.\ 2012; Castellano
et al.\ 2012).

We generate simulated images of each source in all {\it HST},
ground-based, and {\it Spitzer}/IRAC wavelength channels.  Artificial
images of individual sources in the ground-based and {\it
  Spitzer}/IRAC channels are produced by convolving the simulated {\it
  HST} images with the PSF-matching kernels we derive with
\textsc{mophongo} (Labb{\'e} et al.\ 2013).  These images are then
added to sections of the CANDELS UDS, COSMOS, EGS, GOODS-North, and
GOODS-South, and ERS fields, catalogs are constructed, and sources are
selected using exactly the same procedures as we apply to the real
observations.  We include both the criteria used for our pre-selection
and our confirmation criteria (i.e., $P(z>8)>0.9$) in computing the
selection volume.  We implement these criteria in an identical way as
they are applied to the observations.

For example, to be included in our selection volume estimates,
simulated sources are determined to be pre-selected using the criteria
we describe in \S3.2.1 or \S3.3.1.  For simulated sources within the
CANDELS-UDS and CANDELS-COSMOS data sets, this means that their
cumulative probability of lying at $z>8$ must be greater than 50\%,
before the addition of any $Y_{105}$-band data.  In addition,
simulated sources (over the CANDELS-UDS/COSMOS/EGS fields) must show a
probability $>$90\% of lying at $z>8$ after the inclusion of the flux
constraint (0$\pm$12 nJy: nominally the flux constraint one would
obtain for a $z\sim9$-10 galaxy based on a single orbit of
$Y_{105}$-band observations) and have a measured $J_{125}-H_{160}$
color $>$0.5 mag.  Our simulation results make it clear how important
the pre-selection can be.  While increasing the efficiency of our
search results significantly, pre-selection can also introduce a
modest amount of incompleteness into the $z\sim9$-10 samples we
identify from CANDELS, particularly at $z<9$ (where it is $\sim$40\%
from the pre-selection step alone).

In addition to our considering the selection of sources from
CANDELS-UDS, CANDELS-COSMOS, and CANDELS-UDS fields, we also consider
the selection of $z\sim9$ and $z\sim10$ galaxies from the CANDELS
GOODS-North, CANDELS GOODS-South, and ERS fields.

In computing the number of confirmed $z=9$-10 sources from our
program, we assume all the sources in Table~\ref{tab:catalog_conf} are
bona-fide $z\sim9$-10 galaxies and there is no contamination in our
selection.  This would appear to be a good assumption, given that the
typical $z\sim9$-10 candidate formally prefers a $z>8$ solution at
99\% likelihood.  We do not include EGS910-2 and UDS910-1 in our LF
calculation, since they do not meet our formal criteria for inclusion
(but nevertheless appear to be probable $z\gtrsim8.5$ galaxies).  We
suppose that all of the candidates from our follow-up program that
were not explicitly confirmed by that program to lie at $z<8.4$ (all
but one of these candidates was detected at $\geq$2$\sigma$ in the
follow-up $Y_{105}$-band observations and are therefore unlikely
$z>8.4$).  The $z\sim9$ candidate GS-z9-5 is modeled as 0.5 $z\sim9$
galaxy (i.e., $\sim$50\% probability of contamination), given its
computed $P(z>8)$ was only 0.66 (Figure~\ref{fig:sed_south}).  We
ignore the impact of possible lensing amplification on one source in
our selection (EGS910-3) given the size of the magnification factor
(0.25 mag) and the fact that source volume and magnification factor
trade off in such a way to have little impact on the derived LF.

Our $z\sim9$ and $z\sim10$ LF results are presented in
Figure~\ref{fig:complf} and also tabulated in Table~\ref{tab:swlf}.
In computing the uncertainties on the $z\sim9$ and $z\sim10$ $UV$ LFs,
we also include the expected large-scale structure uncertainties,
using the results from the cosmic variance calculator of Trenti \&
Stiavelli (2008) and the observed comoving volume density.  For
context, the earlier LF results of McLure et al.\ (2013), Oesch et
al.\ (2013, 2014), Bouwens et al.\ (2014b, 2015), and McLeod et
al.\ (2015) are presented in Figure~\ref{fig:complf}.

Given the limited numbers of $z\sim9$-10 candidates in our samples and
some arbitrariness in the choice of bin centers (and width) for our
stepwise $z\sim9$-10 LF results, it is conceivable that our
$z\sim9$-10 LF results could depend on how we bin our sample.  We
therefore also model the bright end of the LF as a power law
(motivated by the results of e.g. Bowler et al.\ 2014 and Bouwens et
al.\ 2015).  By marginalizing over both the normalization and
power-law slope of the model LFs, we derive constraints (68\%
confidence levels) on the volume density of $z\sim9$-10 galaxy
candidate for a given $M_{UV}$.  These results are presented in
Figure~\ref{fig:complf} and shown in Table~\ref{tab:modellf}.

\begin{figure}
\epsscale{1.18} \plotone{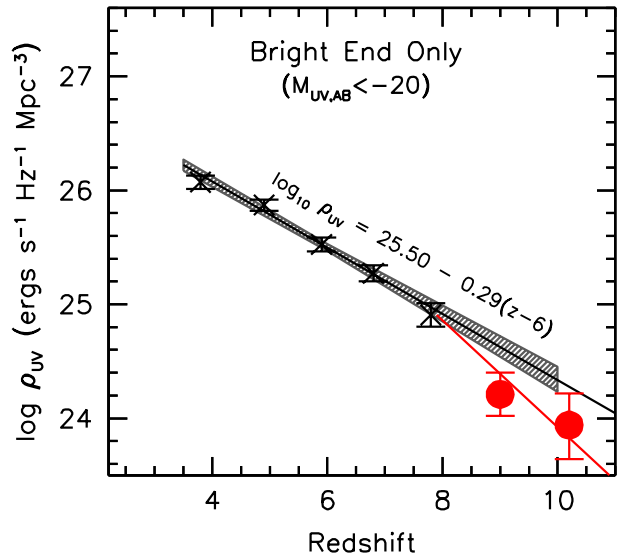}
\caption{The present determinations of the $UV$ luminosity density of
  galaxies at $z\sim9$ and $z\sim10$ (\textit{solid red circles})
  brightward of $-20$ mag using the present search over all 5 CANDELS
  fields.  The black crosses show the luminosity density of galaxies
  brightward of $-20$ mag determined by Bouwens et al.\ (2015) at
  $z\sim4$, $z\sim5$, $z\sim6$, $z\sim7$, and $z\sim8$.  The shaded
  region shows the evolutionary trend to $z>8$ in the $UV$ luminosity
  density preferred at 68\% confidence.  The red line shows the
  observed evolution in the $UV$ luminosity density beyond $z\sim8$
  and can be approximately represented by $d\log_{10} \rho_{UV}/dz
  =-0.45\pm0.07$ dex.  The $UV$ luminosity density we find for
  luminous $z\sim9$ and $z\sim10$ galaxies lies below a simple
  extrapolation of the evolution from $z\sim4$ to
  $z\sim8$.\label{fig:ldevol}}
\end{figure}

To guide expectations, we also included on Figure~\ref{fig:complf} the
LFs we derive extrapolating the LF results from Bouwens et al.\ (2015)
to $z\sim9$ and $z\sim10.2$ (the mean redshift of our $z\sim9$ and
$z\sim10$ derived from our selection volume simulations).  In deriving
a fitting formula from the Bouwens et al.\ (2015) results, we only
consider the evolution over the range $z\sim8$ to $z\sim5$ where the
evolution of the $UV$ LF evolves in a relatively smooth manner.  This
results in the Schechter parameters depending on redshift in the
following manner: $M_{UV} ^{*} = (-20.97\pm0.10) + (0.17\pm0.10) (z -
6)$, $\phi^* = (0.45_{-0.08}^{+0.10}) 10^{(-0.21\pm0.09)(z-6)}10^{-3}
\textrm{Mpc}^{-3}$, $\alpha = (-1.91\pm0.05) + (-0.13\pm0.05)(z-6)$.
This fitting formula implies a $M^*=-20.45$, $\phi^* =
0.10\times10^{-3}$ Mpc$^{-3}$, and $\alpha=-2.3$ at $z\sim9$ and
$M^*=-20.28$, $\phi^* = 0.059\times10^{-3}$ Mpc$^{-3}$, and
$\alpha=-2.46$ at $z\sim10.2$.  The best-fit evolutionary scenario
from Bowler et al.\ (2014) is very similar to what we present above
(where it was found that $dM^*/dz\sim0.2$).

Relative to these extrapolations of $z=4$-8 results to $z>8$, our
present LFs are typically 1.5$\times$ lower in the mean, consistent
with a slightly faster evolution at $z>8$.  Results from the above
fitting formula are also featured on Figure~\ref{fig:lfall} where we
combine the new LF results with previous results from the literature
at $z\sim4$, $z\sim5$, $z\sim6$, $z\sim7$, and $z\sim8$.

The present estimates of the volume density of particularly bright
$z\sim9$ galaxies are the first available in the literature and have
comparable uncertainties to the bright end of the present LF
determinations at $z\sim10$ (and as derived earlier: Oesch et
al.\ 2014; Bouwens et al.\ 2015).

\subsection{Evolution of the $UV$ Luminosity Density for Luminous Galaxies from $z\sim10$ to $z\sim4$}

With the present constraints on the volume density of bright galaxies
at $z\sim9$ and $z\sim10$, we can now examine the evolution of the
luminosity density of galaxies in the rest-frame $UV$ with cosmic
time.

Integrating up the light in our binned representation of the bright
end of the $z\sim9$ and $z\sim10$ LFs to $-20$ mag (approximately the
faint-end limit of the present bright search), we derive a total $UV$
luminosity density $\rho_{UV}$ of $10^{24.21_{-0.19}^{+0.13}}$
erg/s/Hz/Mpc$^3$ at $z\sim9$ and $10^{23.94_{-0.28}^{+0.18}}$
erg/s/Hz/Mpc$^3$ at $z\sim10$.  These inferred luminosity densities
are 5$_{-2}^{+3}\times$ and 8$_{-3}^{+9}\times$ lower, respectively,
than Bouwens et al.\ (2015) found at $z\sim8$.

In Figure~\ref{fig:ldevol}, we compare the derived luminosity density
$\rho_{UV}$ at $z\sim9$ and $z\sim10$ with the $UV$ luminosity density
Bouwens et al.\ (2015) derived from $z\sim4$ to $z\sim8$ to the same
limiting magnitude.  Also shown on this figure is the best-fit
evolution (68\% confidence intervals) derived for the $UV$ luminosity
density based on the $z\sim4$-8 results.  Our new luminosity density
results $\rho_{UV}$ at $z\sim9$ and $z\sim10$ are
$2.6_{-0.9}^{+1.5}\times$ and $2.2_{-1.1}^{+2.0}\times$ lower than the
extrapolated luminosity densities at $z\sim9.0$ and $z\sim10.2$.

The present results are therefore consistent with a more rapid
evolution at $z>8$ than between $z\sim8$ and $z\sim4$, as first noted
by Oesch et al.\ (2012), with a best-fit $d\log_{10} \rho_{UV} /dz$
evolution of $-0.45\pm0.07$ vs. the $-0.29\pm0.02$ evolution observed
from $z\sim8$ to $z\sim4$.  Interestingly enough, the evolution we
derive here is completely consistent with the $d\log_{10}
\rho_{UV}/dz=-0.54_{-0.36}^{+0.19}$ scaling that Oesch et al.\ (2012)
previously derived.  It is also consistent with the $(1+z)^{-10.9}$
evolutionary scalings considered in Oesch et al.\ (2014: see also
Oesch et al.\ 2013).

Other authors (Ishigaki et al.\ 2015; Oesch et al.\ 2015; McLeod et
al.\ 2015) have remarked that there is less evidence for a faster
evolution in the $UV$ luminosity density at $z>8$ integrated to lower
luminosities, particularly when incorporating new results from the
full Hubble Frontier Fields program (Lotz et al.\ 2014; Coe et
al.\ 2015).  If the HUDF/XDF is systematically underdense in
$z\sim9$-10 galaxies or if the evolution of the LF proceeds more
smoothly for lower luminosity galaxies, one might expect such a
result.  A much more definitive exploration of this issue will be
possible by considering the much larger number of faint sources at
$z\sim9$-10 expected from the full Frontier Fields program (Coe et
al.\ 2015).

\section{Summary}

In this paper, we provide improved constraints on the volume density
of especially luminous ($H_{160,AB}<27$) $z\sim9$ galaxies by making
use of a search over all 5 CANDELS fields (750 arcmin$^2$ area in
total).  We also rederive the bright end of the $z\sim10$ LF and
extend it slightly fainter taking advantage of the additional search
power provided by the S-CANDELS data (Ashby et al.\ 2015) over the
CANDELS-UDS, COSMOS, and EGS fields.

To obtain these constraints, we extend the earlier sample of 6 bright
$z\sim9$ and $z\sim10$ galaxies identified by Oesch et al.\ (2015) to
make use of the search area over all 5 CANDELS fields, including an
additional $\sim450$ arcmin$^2$ area over the CANDELS-UDS,
CANDELS-COSMOS, and CANDELS-EGS fields.  We also expand the Oesch et
al.\ (2014) search over the GOODS-South and GOODS-North to obtain a
more complete selection of galaxies over the redshift range
$z\sim8.4$-9.0 (see \S5).

To identify a robust sample of bright $z\sim9$ and $z\sim10$ galaxies
over this new area, we employed a two-part strategy for selecting
sources.  To begin, we selected those sources that showed the highest
probability of corresponding to $z\sim9$ and $z\sim10$ galaxies based
on the existing observations.  Such identifications were possible,
given the depth of the {\it HST} and ground-based imaging observations which
placed constraints on both the sharpness of a possible spectral break
at $1.2\mu$m and also the spectral slope blueward of the break.

The second step was to obtain deep observations on each of these
candidates at $1.05\mu$m to test the nature of these candidates.
These follow-up observations were obtained with the 11-orbit {\it HST}
program z9-CANDELS (Bouwens 2014: GO 13792).

Using the new {\it HST} observations from the z9-CANDELS program, we find
that 5 out of the 12 $z\sim9$-10 candidates we have followed up with
our program appear to be bona-fide bright $z\sim9$-10 galaxies
(Figure~\ref{fig:sed_conf}).

Combining our new samples with previous samples of luminous
$z\sim9$-10 candidates over CANDELS GOODS-North and GOODS-South (Oesch
et al.\ 2014) and also including four probable $z\sim8.4$-9.0 galaxies
we identified over the GOODS-S field (finding no additional bright
sources over the GOODS-N field), we identify a total sample of 15
bright $z\sim9$-10 candidates over the five CANDELS fields
(Table~\ref{tab:catalog_conf}).  This is $\gtrsim$2$\times$ larger
than the sample of luminous $z\sim9$-10 galaxies we had identified
earlier in Oesch et al.\ (2014).

We use these larger samples of $z\sim9$-10 galaxies to derive improved
constraints on the bright end of the $z\sim9$ and $z\sim10$ LFs.  We
achieve these results by simulating both stages in the selection
process (for the CANDELS-UDS, COSMOS, and EGS fields) -- or a single
stage in the case of the CANDELS GOODS-North, GOODS-South, or ERS
fields -- to arrive at statistically-robust conclusions.

As one would expect with any follow-up program of limited size, the
z9-CANDELS program is not able to observe all potential $z\sim9$ and
$z\sim10$ galaxies over the CANDELS-UDS, COSMOS, and EGS
fields\footnote{Several possible examples of missed $z\sim9$-10
  candidates could include those lower-probability candidates
  tabulated in Appendix C.} and therefore likely suffers from moderate
incompleteness, particularly for galaxies at $z<9$, due to our
exclusively preselecting for follow-up sources with redder
$J_{125}-H_{160}$ $>$0.5 colors than galaxies which exist in this
range.  Some incompleteness would also result from the detection of
spurious flux in the optical bands ($\sim$25\% effect).  Nevertheless,
we remark that all such effects are included in our selection volume
simulations and our focusing on $J_{125}-H_{160}>0.5$ galaxies means
our search is most complete at $z>9$.

We would expect these results to be improved, particularly towards
fainter magnitudes, if even deeper observations over the CANDELS-UDS,
CANDELS-COSMOS, and CANDELS-EGS fields could be obtained.  Deeper
observations in the F105W band should increase the size of our bright
($H_{160,AB}\lesssim 27$) $z\sim9$-10 samples over the CANDELS-WIDE
fields by improving the completeness of our search results at
$z<9$.\footnote{While our selection of $z\sim9$-10 galaxies is fairly
  complete at $z>9$, it suffers greater incompleteness at
  $z\sim8.4$-9.0, due to our use of a $J_{125}-H_{160}>0.5$ criterion
  in preselecting candidate $z\sim9$-10 galaxies to follow up.  The
  addition of $Y_{105}$-band observations to these fields would allow
  such sources to be selected much more efficiently.}  Meanwhile,
deeper observations in various redder bands (e.g., $JH_{140,AB}$)
would allow us to extend these searches to even fainter magnitudes
(i.e., to $>$26.5 mag).  We would also expect gains from analysis of
the 500-orbit, 500 arcmin$^2$ BoRG$_{[z910]}$ program (Trenti 2014).

\acknowledgements

We thank Jim Dunlop and Steve Finkelstein for valuable conversations.
We thank Steve Willner for providing us with feedback on an earlier
draft of this manuscript.  This work is based in part on observations
made with the {\it Spitzer} Space Telescope, which is operated by the
Jet Propulsion Laboratory, California Institute of Technology under a
contract with NASA.  We acknowledge the support of NASA grant
NAG5-7697, NASA grant {\it HST}-GO-11563, and a NWO vrij competitie
grant 600.065.140.11N211.

\appendix

\begin{figure*}
\epsscale{0.87} \plotone{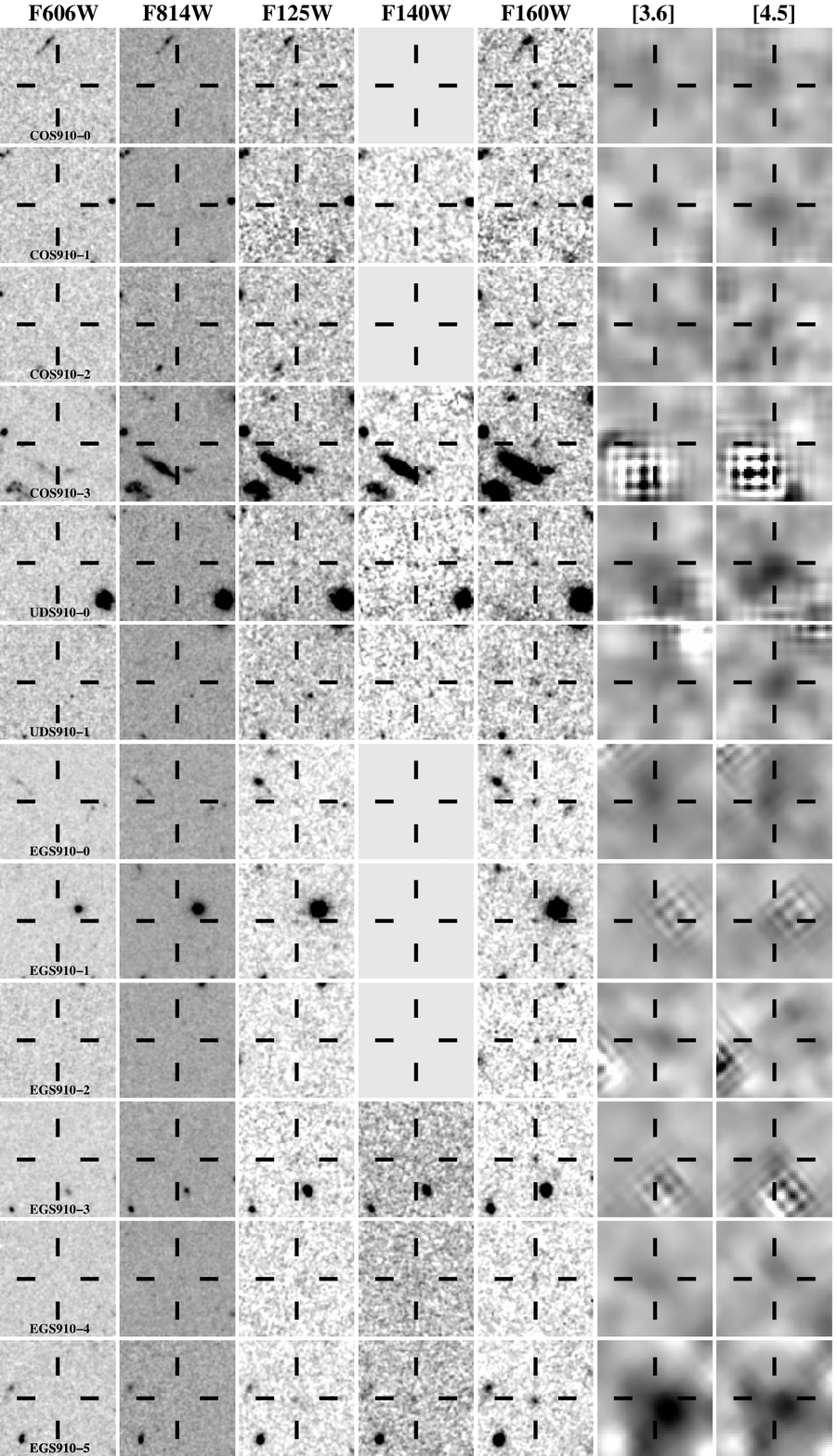}
\caption{{\it HST} + {\it Spitzer}/IRAC images ($6''\times6''$) of all
  preselected candidate $z\sim9$-10 galaxies over the CANDELS-UDS +
  CANDELS-COSMOS + CANDELS-EGS fields that were preselected for
  targeted follow-up with the z9-CANDELS program.  The best-fit model
  flux from neighboring sources has been removed in the {\it
    Spitzer}/IRAC images shown here.  Our preselected CANDELS-UDS and
  CANDELS-COSMOS candidates (1) are estimated to show a $>$50\%
  probability of corresponding to a $z>8$ galaxy and (2) can be
  confirmed to lie at $z>8$ with $>$90\% confidence with the addition
  of a single orbit of {\it HST} follow-up observations (assuming a
  flat redshift prior and the EAZY SED template set).  Each of the
  preselected CANDELS-EGS candidates shown here can be confirmed to be
  secure at $>$90\% probability with a single orbit of {\it HST}
  follow-up observations.  Each of these candidates was subject to
  1-orbit follow-up observations with our z9-CANDELS program at
  $1.05\mu$m.\label{fig:stamp1}}
\end{figure*}

\begin{figure*}
\epsscale{1.18} \plotone{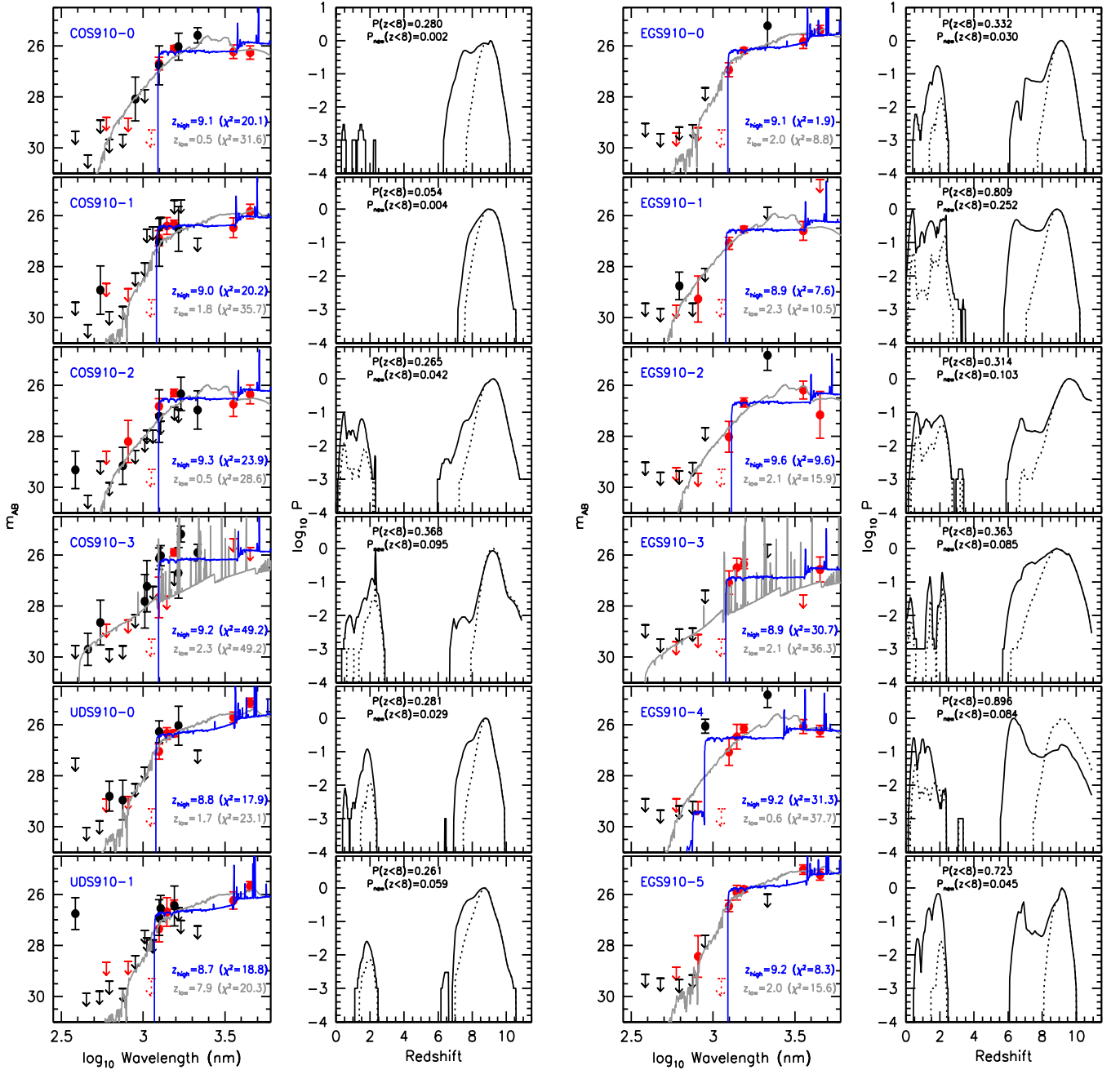}
\caption{(\textit{leftmost column}) Best-fit SED models to the
  observed {\it HST}+{\it Spitzer}/IRAC+ground-based photometry of the preselected
  candidate $z\sim9$-10 galaxies we have identified for targeted
  follow-up with the z9-CANDELS follow-up program.  The dotted red
  upper limits show the approximate constraints we will be able to set
  on the $1.05\mu$m fluxes of the candidates, assuming they are at
  $z\gtrsim8.4$.  The points and lines are otherwise as in
  Figure~\ref{fig:sed_conf}.  (\textit{second leftmost column})
  Redshift likelihood distribution for the same 12 $z\sim9$-10
  candidates using current observations (\textit{solid lines}) and
  also making use of our 1-orbit $Y_{105}$-band follow-up observations
  (\textit{dotted lines}) assuming the sources are at $z\gtrsim9$.
  $P(z<8)$ and $P_{new}(z<8)$ indicate the probability that our
  candidates are at $z<8$ using the current observations and including
  our follow-up observations (again assuming they are $z\gtrsim9$).
  It should be clear that our z9-CANDELS follow-up observations should
  significantly improve our confidence in the present set of
  $z\gtrsim9$ candidates, increasing it from 72-95\% to 95.8-99.8\%.
  \label{fig:sed1}}
\end{figure*}

\begin{figure*}
\epsscale{0.94} \plotone{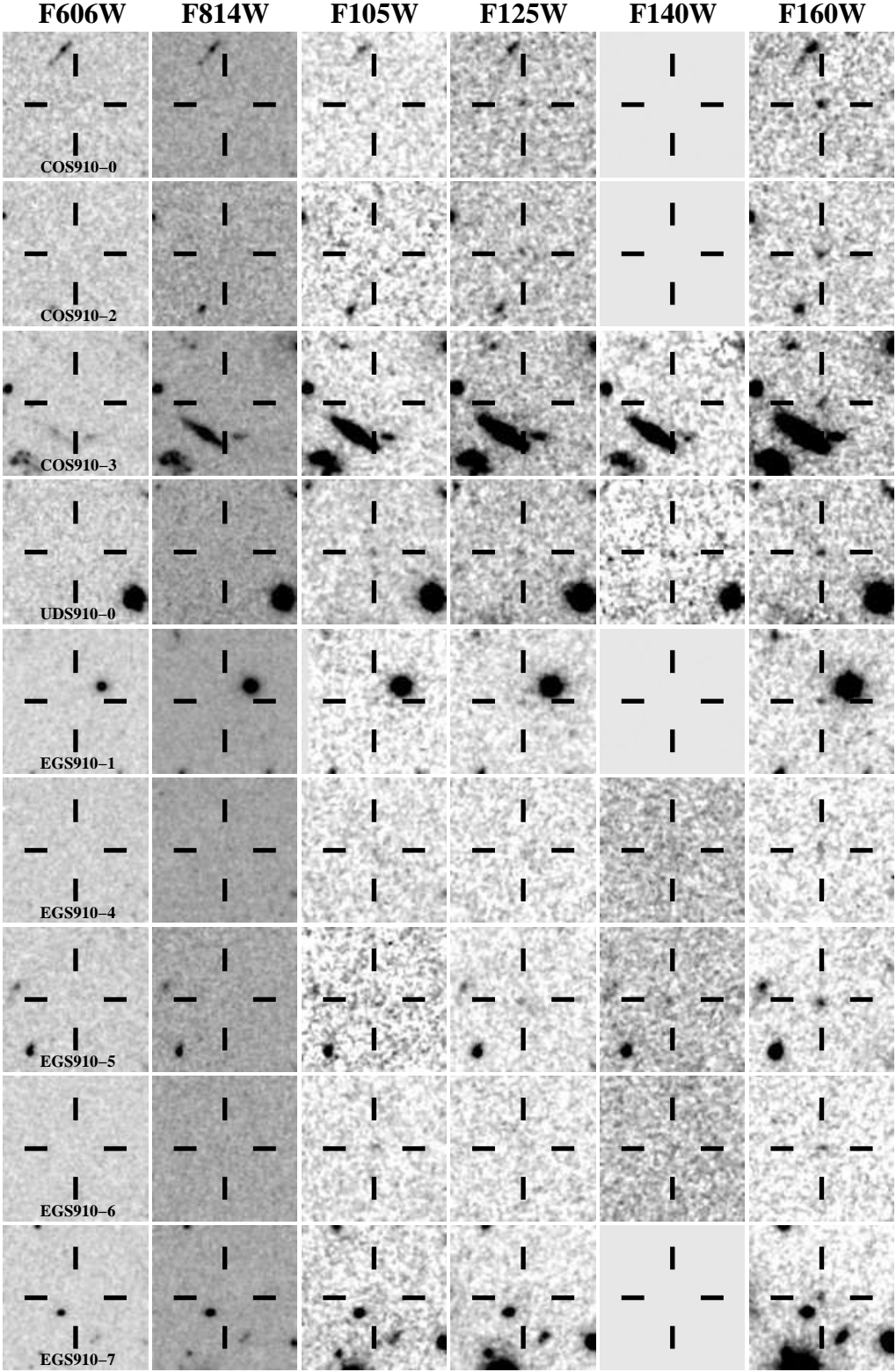}
\caption{{\it HST} + {\it Spitzer}/IRAC images for 7 targeted
  candidate $z\sim9$-10 galaxies which follow-up observations from our
  {\it HST} z9-CANDELS program failed to confirm as probable $z\geq9$
  galaxies using {\it HST} follow-up observations with our z9-CANDELS
  program.  F105W-band observations were also obtained for the sources
  shown in the lowest 2 rows of this figure from the z9-CANDELS {\it
    HST} program.  While neither source was not explicitly targeted
  for observations by our program due to a relatively low prior
  probability of being at $z>8$, we could not rule out that
  possibility and so took advantage of the new data to gain more
  insight into their nature.\label{fig:stamp_notconf}}
\end{figure*}

\begin{figure*}
\epsscale{1.18} \plotone{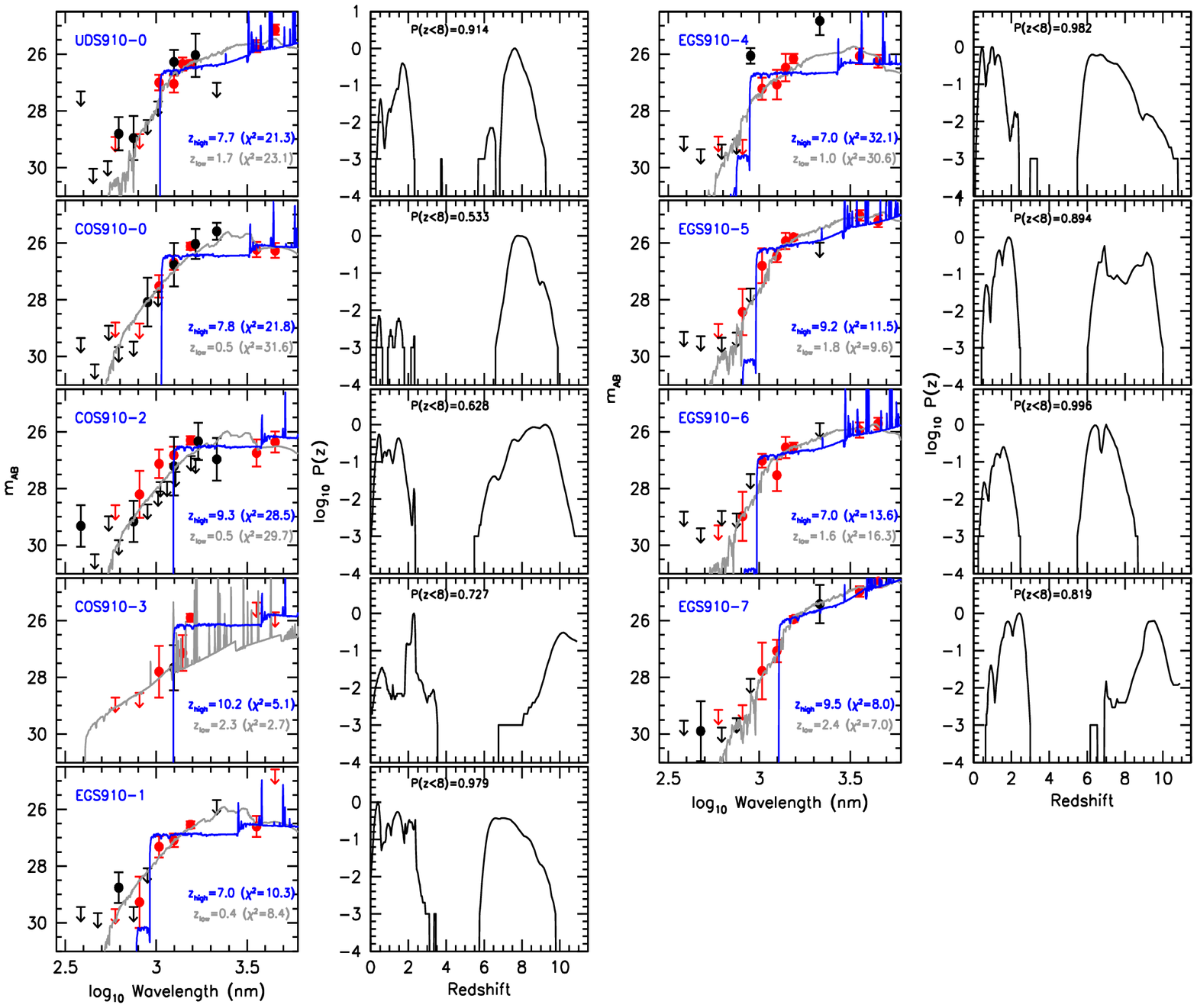}
\caption{(\textit{left}) Best-fit SED models to the observed {\it
    HST}+{\it Spitzer}/IRAC+ground-based photometry for 7 targeted
  candidate $z\sim9$-10 galaxies (COS910-0, COS910-2, COS910-3,
  EGS910-1, EGS910-4, EGS910-5) which were not confirmed as probable
  $z>8.4$ galaxies using follow-up observations from the z9-CANDELS
  program.  The points and lines are otherwise as in
  Figure~\ref{fig:sed_conf}.  F105W-band observations were also
  obtained for the sources shown in the lowest 2 rows of this figure
  (EGS910-6 and EGS910-7).  While neither source was not explicitly
  targeted for observations by our program due to a relatively low
  prior probability of being at $z>8$, we could not rule out that
  possibility and so took advantage of the new data to gain more
  insight into their nature.  (\textit{right}) Redshift likelihood
  distribution for these $z\sim9$-10 candidates incorporating both our
  follow-up observations with {\it HST} and the {\it HST}+{\it
    Spitzer}/IRAC+ground-based observations that were used in the
  pre-selection (\textit{solid lines}).\label{fig:sed_notconf}}
\end{figure*}

\section{A.  Postage Stamps and SEDs of the Targeted Candidates}

We identified 12 candidate $z=9$-10 sources over the CANDELS-UDS,
CANDELS-COSMOS, and CANDELS-EGS fields that we explicitly targeted for
follow-up observations.

Postage stamp images of these 12 candidates are presented in
Figure~\ref{fig:stamp1}.  As should be obvious from this figure, all
six of the present $z\sim9$-10 candidates show clear detections in the
$H_{160}$-band, as well as significant $\sim2$-3$\sigma$ detections in
the $J_{125}$-band and $JH_{140}$-band observations (where available),
as well as in the S-CANDELS {\it Spitzer}/IRAC data.

The observed photometry to those candidate sources are presented in
Figure \ref{fig:sed1}, along with SED fits to a model $z>8$ galaxy and
a model $z<3$ galaxy.  Also shown on this figure is the redshift
likelihood distribution (\textit{solid black line}) based on the
photometry we have available for each candidate in the $\sim$20
different wavelength channels ({\it HST} + {\it Spitzer}/IRAC +
ground-based observations).  In addition, this figure presents the
redshift likelihood distribution we would expect, assuming these
candidates are not detected in the single orbit of follow-up
$Y_{105}$-band observations from the z9-CANDELS program.

All 12 candidates show similar $H_{160}-[3.6]$ colors (see also
Wilkins et al.\ 2016), red $[3.6]-[4.5]$ colors, and observed sizes
(Holwerda et al.\ 2015; Shibuya et al.\ 2015) to the first samples of
particularly luminous $z\sim9$-10 galaxies identified by Oesch et
al.\ (2014).

\section{B.  Targeted $z\sim9$-10 Candidates that are most likely at $z<8.4$}

A fraction of the candidate $z\sim9$-10 galaxies targeted by our
follow-up observations from the z9-CANDELS program appear not to be
at $z>8$.  In Figure~\ref{fig:stamp_notconf}, we present postage
stamps for 9 such sources (7 explicitly targeted by our program and 2
lower probability $z\sim9$-10 candidates which were incidentally
targeted).  In almost all of the candidates which are not confirmed by
our follow-up observations, the sources show a $\geq2\sigma$ detection
in the $Y_{105}$-band.  Figure~\ref{fig:sed_notconf} shows the
best-fit SED models for the tentative $z\sim9$-10 galaxies that were
not confirmed by observations from our follow-up program.

Detailed comments on specific candidate $z\sim9$-10 galaxy that were
targeted for follow-up with our z9-CANDELS program can be found
below:

\noindent \textbf{COS910-0:} Follow-up of COS910-0 in the
$Y_{105}$-band shows a $3\sigma$ detection at $1.05\mu$m.  A detailed
fit to its SED suggests that it is actually a star-forming galaxy at
$z=7.8$.  Its inclusion in our original sample of $z>8.4$ candidate
galaxies occurred due to its measured $J_{125}-H_{160}$ color, which
was likely redder than reality due to the impact of noise.  \\
\noindent \textbf{COS910-2:} COS910-2 is detected at $2.2\sigma$ in
the $Y_{105}$-band follow-up observations.  Such a detection is not
expected if the galaxy is actually at $z\gtrsim9$, and so the source
must have a redshift of $z\lesssim8$.  It may correspond to either a
$z\sim0.5$ or a $z\sim8$ galaxy.\\
\noindent \textbf{COS910-3:} Follow-up of COS910-2 in the
$Y_{105}$-band and the $JH_{140}$-band (1/3 of an orbit) shows a
$1\sigma$ detection at $1.05\mu$m and $2\sigma$ detection in the
$JH_{140}$ band.  Perhaps, most importantly, the deeper photometry
over the source in the $JH_{140}$ band confirms that the source has a
$JH_{140}-H_{160}$ color of 1.25 mag.  While this is consistent with
the source having a redshift of the $z\gtrsim11$, one would not expect
to detect the source at $2\sigma$ in the $J_{125}$ band in this case.
These results suggest that the apparent break in the spectrum at
$1.5\mu$m is not especially sharp and this source shows faint but
detectable flux to much bluer wavelengths.\\
\noindent \textbf{UDS910-0:} Follow-up of UDS910-0 in the $Y_{105}$
band shows a $4\sigma$ detection at $1.05\mu$m.  Therefore, this
source cannot plausibly correspond to a $z\gtrsim8.4$ galaxy.  The
best-fit redshift we compute for the source is $z\sim7$.  Like
COS910-0, its inclusion in our sample of $z\sim9$-10 galaxies likely
occurred as a result of noise in the measured $J_{125}-H_{160}$
color.\\
\noindent \textbf{EGS910-1:} Follow-up of EGS910-1 in the
$Y_{105}$-band shows a $3\sigma$ detection at $1.05\mu$m, consistent
with a redshift of $z<8$.\\
\noindent \textbf{EGS910-4:} Follow-up of EGS910-4 in the
$Y_{105}$-band shows a $3\sigma$ detection at $1.05\mu$m, consistent with a redshift of $z<8$.\\
\noindent \textbf{EGS910-5:} Follow-up of EGS910-5 in the
$Y_{105}$-band shows a $1.7\sigma$ detection at $1.05\mu$m, strongly
suggesting it is not at $z>8$.\\
\noindent \textbf{EGS910-6:} EGS910-6 is detected at 4.5$\sigma$ in
our $Y_{105}$-band follow-up observations at $1.05\mu$m, providing
clear evidence it is at $z<8$.  A $z<8$ solution is preferred at 99.8\%
confidence (Figure~\ref{fig:sed_notconf}).\\
\noindent \textbf{EGS910-7:} Follow-up of EGS910-7 in the
$Y_{105}$-band only shows a $1.1\sigma$ detection at $1.05\mu$m.
However, the $H_{160}-[3.6]$ color is sufficiently red that it seems
more consistent with a lower-redshift galaxy than a $z>8$ galaxy.  The
nature of this source is unclear.\\

\section{C.  Other Candidate $z\sim9$-10 Galaxies over the CANDELS-UDS, COSMOS, and EGS Fields}

When putting together a compilation of the most promising $z\sim9$-10
candidates to target with follow-up {\it HST} observations, we experimented
with a variety of different procedures to identify possible
$z\sim9$-10 galaxy candidates.  As a result, we identified a larger
number of possible $z=9$-10 galaxy candidates than we could thoroughly
follow up with the 11 orbits allocated to our {\it HST} program.

While the redshift likelihood distributions we derived for these
candidates suggested that the vast majority of them likely
corresponded to sources at $z<8$, it is nevertheless possible that
several of these candidates might have redshifts in excess of
$z\sim8$.

\begin{figure}
\epsscale{0.8} \plotone{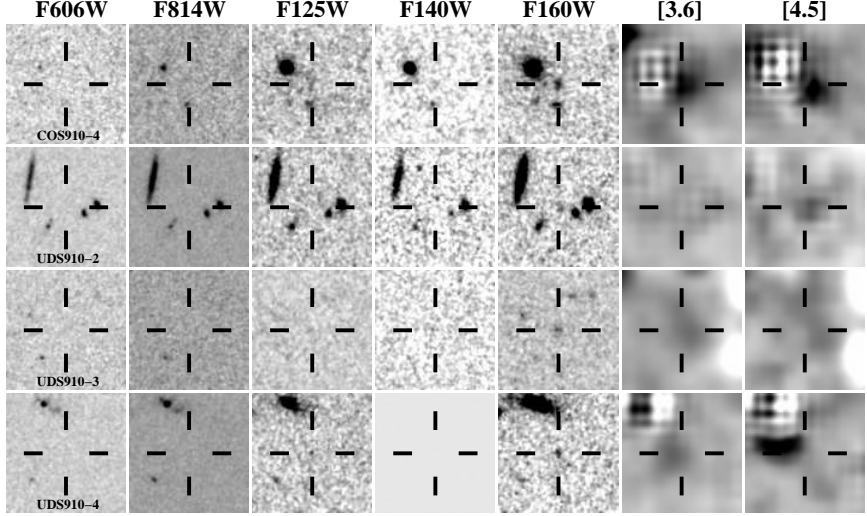}
\caption{{\it HST} + {\it Spitzer}/IRAC images ($6''\times6''$) of 4 possible
  $z\sim9$-10 candidate galaxies that did not meet our criteria for
  preselection and hence were not targeted by our z9-CANDELS
  program.  Sources are shown in the same
  passbands, as in Figure~\ref{fig:stamp1}.  None of these sources is
  explicitly targeted with our z9-CANDELS follow-up
  program.\label{fig:stamp_low}}
\end{figure}

\begin{figure}
\epsscale{1.} \plotone{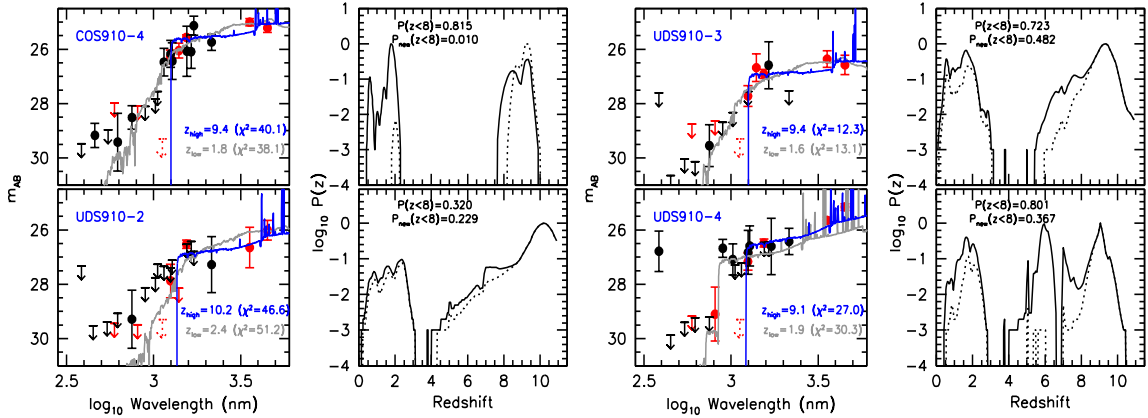}
\caption{(\textit{left}) Best-fit SED models to the observed
  {\it HST}+{\it Spitzer}/IRAC+ground-based photometry of 4 possible $z\sim9$-10
  candidate galxies that did not meet our criteria for preselection
  and hence were not targeted by our z9-CANDELS follow-up program.
  Symbols are as in Figure~\ref{fig:sed_conf} and \ref{fig:sed1}.
  (\textit{right}) Redshift likelihood distribution for these 6
  $z\sim9$-10 candidates.\label{fig:sed_low}}
\end{figure}

We provide a short list of the ``lower-probability'' $z\sim9$-10
galaxy candidates we identified over the CANDELS COSMOS, UDS, and EGS
fields in Table~\ref{tab:catalogex}.  These sources did not meet our
criteria for preselection, but nevertheless may correspond to
$z\sim9$-10 galaxies.  Figure~\ref{fig:stamp_low} shows postage stamp
cut-outs of these sources, while Figure~\ref{fig:sed_low} illustrates
fits to their spectral energy distributions and our derived redshift
likelihood contours for these sources.

\begin{deluxetable*}{cccccc}
\tablewidth{0pt} \tablecolumns{6} \tabletypesize{\footnotesize}
\tablecaption{Possible $z\sim9$-10 Galaxies over the CANDELS UDS,  COSMOS, and EGS program that did not satisfy for our criteria for preselection and hence were not targeted for follow-up observations.\label{tab:catalogex}} \tablehead{
  \colhead{ID} & \colhead{R.A.} & \colhead{Dec} &
  \colhead{$H_{160,AB}$} & \colhead{$z_{phot}$} & \colhead{P($z>8$)\tablenotemark{a}}}
\startdata 
COS910-4 & 10:00:15.52 & 02:17:01.5 & 25.6$\pm$0.1 & 9.3 & 0.18 \\ 
UDS910-2\tablenotemark{b} & 02:17:13.08 & $-$05:15:55.4 & 26.6$\pm$0.2 & 10.2 & 0.68 \\ 
UDS910-3 & 02:17:52.38 & $-$05:15:06.3 & 26.9$\pm$0.2 & 9.4 & 0.28 \\ 
UDS910-4 & 02:17:14.61 & $-$05:15:15.7 & 26.6$\pm$0.2 & 9.1 & 0.50
\enddata
\tablenotetext{a}{Redshift likelihood is computed using the flux measurements for these sources in all photometric bands shown in Table~\ref{tab:dataset}.}
\tablenotetext{b}{While the source UDS910-2 nominally has $>$50\% probability of lying at $z>8$, it was not targeted with our follow-up program z9-CANDELS because it could not be confirmed to be $>$90\% probability candidate with the addition of a single orbit of {\it HST} observations.}
\end{deluxetable*}

\end{document}